\documentclass[11pt]{article}

\usepackage{latexsym,amsmath,epsfig}

\DeclareFontFamily{OT1}{rsfs10}{}
\DeclareFontShape{OT1}{rsfs10}{m}{n}{ <-> rsfs10 }{}
\DeclareMathAlphabet{\mathscript}{OT1}{rsfs10}{m}{n}

\numberwithin{equation}{section}

\newcommand{\ns}{\normalsize}

\newcommand{\tr}{\mathrm{tr}}
\newcommand{\HdR}{H_{\mathrm{DR}}}
\newcommand{\orb}{S^1/Z_2}

\newcommand{\hf}{\frac{1}{2}}

\newcommand{\w}{\wedge}
\newcommand{\ts}{\times}

\newcommand{\genus}[1]{\text{genus}\,(#1)}

\newcommand{\CC}{\mathbf{C}}
\newcommand{\ZZ}{\mathbf{Z}}
\newcommand{\RR}{\mathbf{R}}
\newcommand{\PP}{\mathbf{P}}
\newcommand{\PPa}{\PP^1_a}

\newcommand{\ab}{{\bar a}}
\newcommand{\bb}{{\bar b}}

\newcommand{\cE}{\mathcal{E}}

\newcommand{\cM}{\mathcal{M}}
\newcommand{\cN}{\mathcal{N}}

\newcommand{\cS}{\mathcal{S}}

\def\d{\delta}

\def\l{\lambda}

\def\p{\pi}

\def\r{\rho}
\def\s{\sigma}

\def\D{\Delta}

\def\G{\Gamma}

\def\L{\Lambda}
\def\O{\Omega}

\def\S{\Sigma}

\begin{document}


\begin{titlepage}

\vspace{-5cm}

\title{
   \hfill{\ns UPR-843T, PUPT-1856} \\[3em]
   {\LARGE Moduli Spaces of Fivebranes on Elliptic Calabi-Yau Threefolds}
       \\[1em] } 
\author
   {\ns Ron Donagi$^1$, Burt A.~Ovrut$^2$ 
      and Daniel Waldram$^3$ \\[0.5em]
   {\it\ns $^1$Department of Mathematics, 
      University of Pennsylvania} \\[-0.3em]
   {\it\ns Philadelphia, PA 19104--6395, USA}\\ 
   {\it\ns $^2$Department of Physics, University of Pennsylvania} \\[-0.3em]
   {\it\ns Philadelphia, PA 19104--6396, USA}\\ 
   {\it\ns $^3$Department of Physics, Joseph Henry Laboratories,}\\[-0.3em]
   {\it\ns Princeton University, Princeton, NJ 08544, USA}}
\date{}

\maketitle

\begin{abstract}

We present a general method for calculating the moduli spaces of
fivebranes wrapped on holomorphic curves in elliptically fibered
Calabi--Yau threefolds, in particular, in the context of heterotic
M~theory. The cases of fivebranes wrapped purely on a fiber curve,
purely on a curve in the base and, generically, on a curve with
components both in the fiber and the base are each discussed in
detail. The number of irreducible components of the fivebrane and
their properties, such as their intersections and phase transitions in
moduli space, follow from the analysis. Even though generic curves
have a large number of moduli, we show that there are isolated curves
that have no moduli associated with the Calabi--Yau threefold. We
present several explicit examples, including cases which correspond to
potentially realistic three family models with grand unified gauge
group $SU(5)$. 

\end{abstract}

\thispagestyle{empty}

\end{titlepage}


\section{Introduction}

If string theory is to be a model of supersymmetric particle physics,
one should be able to find four-dimensional vacua preserving $\cN=1$
supersymmetry. The first examples of such backgrounds arose as
geometrical vacua, from the compactification of the heterotic string
on a Calabi--Yau threefold. Recent advances in string duality have
provided new ways of obtaining geometrical $\cN=1$ vacua, by
compactifying other limits of string theory and, in particular, by
including branes in the background. 

Of particular interest is the strongly coupled limit of the $E_8\ts
E_8$ heterotic string. The low-energy effective theory is
eleven-dimensional supergravity compactified on an $S^1/Z_2$ orbifold
interval, with a set of $E_8$ gauge fields on each of the
ten-dimensional orbifold fixed planes~\cite{hw1,hw2}. To construct a
theory with $\cN=1$ supersymmetry one further compactifies on a
Calabi--Yau threefold $X$~\cite{w:cy}. One is then free to choose
general $E_8$ bundles which satisfy the hermitian Yang--Mills
equations. Furthermore, one can include some number of five-branes. The
requirements of four-dimensional Lorentz invariance and supersymmetry
mean that these branes must span the four-dimensional Minkowski
space, while the remaining two dimensions wrap a holomorphic curve
within the Calabi--Yau threefold~\cite{w:cy,bbs,bvs}. 

In the low-energy four-dimensional effective theory, there is an array
of moduli. There are geometrical moduli describing the dimensions of
the Calabi--Yau manifold and orbifold interval. There are bundle
moduli describing the two $E_8$ gauge bundles. And finally, there are
moduli describing the positions of the fivebranes. It is this last set
of moduli, together with the generic low-energy gauge fields on the
fivebranes, upon which we shall focus in this paper. We note that,
although we consider this problem in the specific case of
M~fivebranes, the moduli spaces are quite general and should have
applications to other supersymmetric brane configurations. 

We shall look at this question in the particular case of an
elliptically fibered Calabi--Yau threefold with section. This expands
on the discussion of we gave in a recent letter~\cite{dlow1}. There we
used this special class of manifolds to construct explicitly an array
of new particle physics vacua. The general structure of the
constructions was given in detail in a second paper~\cite{dlow2}. This
is the companion paper which explains the structure of the five-brane
moduli space as well as the nature of gauge enhancement on the
five-branes. 

The constructions in~\cite{dlow1} and~\cite{dlow2} used the analysis
of gauge bundles on elliptic Calabi--Yau threefolds given by Friedman,
Morgan and Witten~\cite{fmw}, Donagi~\cite{ron} and Bershadsky
\textit{et al.}~\cite{bjps}. The vacua preserved, for example,
$SU(5)$ or $SO(10)$ gauge symmetry with three families. The presence
of the five-branes allowed a much larger class of possible
backgrounds. The number of families is given by an index first
calculated in this context by Andreas~\cite{andreas} and
Curio~\cite{curio}. Curio also gave explicit examples of
bundles where this index was three. Subsequently, the case of
non-simply connected elliptic Calabi--Yau threefolds with bisections
has been considered in~\cite{ack}. We note that, in the M~theory
context, another class of explicit models with non-standard gauge
bundles but with orbifold Calabi--Yau spaces, were first contructed
in~\cite{stie}, while the generic form of the effective four- and
five-dimensional theories, including fivebranes, is given
in~\cite{nse}. 

In constructing these vacua the fivebranes cannot be chosen
arbitrarily~\cite{w:cy}. The boundaries of $\orb$ and the fivebranes
are all magnetic sources for the four-form field strength $G$ of
eleven-dimensional supergravity. The fact that there can be no net
magnetic charge for $G$ in the Calabi--Yau threefold fixes the homology
class of the fivebranes in terms of the gauge bundles and $X$ itself. 
 As discussed in~\cite{dlow1,dlow2}, to describe real fivebranes this
homology class must be ``effective'' in $X$. Mathematically, then, the
problem we wish to solve is to find the moduli space of holomorphic
curves in $X$ in a given effective homology class.

Specific to the M~theory case, we will also include the moduli
corresponding to moving the fivebranes in $\orb$ and, in addition,
their axionic moduli partners~\cite{nse,w:hw,ketal}. These latter
fields are compact scalars which arise as zero modes of the self-dual
three-form $h$ on the fivebranes. The other zero modes of $h$ lead to
low-energy gauge fields. Generically the gauge group is $U(1)^g$,
where $g$ is the genus of the holomorphic
curve~\cite{nse,w:hw,ketal}. Since we will be able to calculate $g$ at
each point in the moduli space, we will also be able to identify the
low-energy gauge multiplets.

One consequence of considering elliptically fibered Calabi--Yau
threefolds with section is that there is a dual F~theory
description~\cite{vafa,vm1,vm2}. For the fivebranes, those wrapping
purely on the fiber of $X$ correspond to threebranes on the F~theory
side~\cite{fmw}. Rajesh~\cite{rajesh} and Diaconescu and
Rajesh~\cite{dr} have recently argued that fivebranes lying 
wholly in the base of $X$ correspond to blow-ups of the corresponding
curve in the F~theory vacuum. We will not comment in detail
on this interesting correspondence. However, we will show that, locally,
the moduli spaces match those expected from duality to F~theory. We
will also comment on how the global structure is encoded on the M~theory
side through a twisting of the axion modulus. This will be discussed
further in~\cite{atwist}. An additional point we will only touch on  
is the structure of additional low-energy fields which can appear when
fivebranes intersect. We will, however, clearly identify
these points in moduli space in our analysis. Finally, we will also
ignore any non-perturbative corrections. In general, since the
low-energy theory is only $\cN=1$, one expects that some of the
directions in moduli space are lifted by non-perturbative
effects, in particular by instantonic membranes stretching between
fivebranes. 

Specifically, we do the following. In section~2, we briefly review the
anomaly cancellation condition in heterotic M-theory, discuss how that
constraint leads to non-perturbative vacua with fivebranes and review
some aspects of homology and cohomology theory required in our
analysis. Properties of elliptically fibered Calabi--Yau threefolds
and a discussion of their algebraic and effective classes are
presented in section~3. Section~4 is devoted to studying the simple
case of the the moduli spaces of fivebranes wrapped purely on the
elliptic fiber. We also comment on global structure of the moduli
space and the relation to F~theory. In section~5, we present two
examples of fivebranes wrapping curves with a component in the
base. We analyze, in detail, the moduli space of these two examples,
including the generic low-energy gauge groups on and possible
intersections of the fivebrane. Techniques developed in sections~4
and~5 are generalized in section~6, where we give a procedure for the
analysis of the moduli spaces of fivebranes wrapped on any holomorphic
curve, generically with both a fiber and a base component. We note a
particular exceptional case which occurs when the fivebrane wraps an
exceptional divisor in the base. Finally, in sections~7,~8 and~9 we
make these methods concrete by presenting three specific examples, two
with a del Pezzo base and one with a Hirzebruch base. Two of these
examples correspond realistic three-family, non-perturbative vacua in
Ho\v rava-Witten theory.


\section{Heterotic M~theory vacua with fivebranes}

\subsection{Conditions for a supersymmetric background}

As we have discussed in the introduction, the standard way to obtain
heterotic vacua in four dimensions with $\cN=1$ supersymmetry, is to
compactify eleven-dimensional M~theory on the manifold $X\ts\orb\ts
M_4$~\cite{hw1,hw2}. Here $X$ is a Calabi--Yau threefold, $S^1/Z_2$ is
an orbifold interval, while $M_4$ is four-dimensional Minkowski
space. This background is not an exact solution, but is good to first
order in an expansion in the eleven-dimensional Planck length. To
match to the low-energy particle physics parameters, the Calabi--Yau
threefold is chosen to be the size of the Grand Unified scale, while
the orbifold is somewhat larger~\cite{w:cy,bd}. In general, there is a
moduli space of different compactifications. There are the familiar
moduli corresponding to varying the complex structure and the K\"ahler
metric on the Calabi--Yau threefold. Similarly, one can vary the size
of the orbifold. These parameters all appear as massless fields in the
low-energy four-dimensional effective action. In general, there are
additional low-energy scalar fields coming from zero-modes of the
eleven-dimensional three-form field $C$. 

The second ingredient required to specify a supersymmetric background
is to choose the gauge bundle on the two orbifold planes. In general,
one can turn on background gauge fields in the compact Calabi--Yau
space. Supersymmetry implies that these fields cannot be
arbitrary. Instead, they are required to satisfy the hermitian
Yang--Mills equations 
\begin{equation}
   F_{ab} = F_{\ab\bb} = 0 \qquad
   g^{a\bb} F_{a\bb} = 0 
\label{hYM}
\end{equation}
Here $a$ and $b$ are holomorphic indices in $X$, $\ab$ and $\bb$ are
antiholomorphic indices, while $g_{a\bb}$ is the K\"ahler metric 
on $X$. Having fixed the topology of the gauge bundle, that is, how the
bundle is patched over the Calabi--Yau manifold, there is then a set
of different solutions to these equations. There are additional
low-energy moduli which interpolate between these different
solutions. In general, the full moduli space of bundles is hard to
analyze. However, when the Calabi--Yau threefold is elliptically
fibered, the generic structure of this moduli space can be calculated
and has been discussed in~\cite{fmw,ron} and also in~\cite{gmod}.

The final ingredient to the background is that one can include
fivebranes~\cite{w:cy,nse}. In order to preserve supersymmetry and
four-dimensional Lorentz invariance, the fivebranes must span the
four-dimensional Minkowski space while the remaining two dimensions
wrap a holomorphic curve within the Calabi--Yau
threefold~\cite{w:cy,bbs,bvs}. In addition, each brane must be parallel
to the orbifold fixed planes. Thus it is localized at a single point
in $\orb$. Again, there are a set of moduli giving the positions of
the five-branes within the Calabi--Yau manifold as well as in the
orbifold interval. As we will discuss below, there are also extra
moduli coming from the self-dual tensor fields on the
fivebranes~\cite{nse,w:hw,ketal}. These fields generically give some
effective $\cN=1$ gauge theory in four-dimensions. Finding the moduli
space of the fivebranes, and some information about the effective
gauge theory which arises on the fivebrane worldvolumes will be the
goal of this paper.  

In summary, the M~theory background is determined by choosing
\begin{itemize}
\item
a spacetime manifold of the form $X\ts\orb \ts M_4$, where $X$ is a
Calabi--Yau threefold
\item
two $E_8$ gauge bundles, $V_1$ and 
$V_2$,  
satisfying the hermitian
Yang--Mills equations~\eqref{hYM} on $X$
\item
a set of fivebranes parallel to the orbifold fixed planes and
wrapped on holomorphic curves within $X$.
\end{itemize}
This ensures that we preserve $\cN=1$ supersymmetry in the low-energy
four-dimensional effective theory.

\subsection{Cohomology condition}

The above conditions are not sufficient to ensure that one has a consistent
background. Anomaly cancellation on both the ten-dimensional orbifold
fixed planes and the six-dimensional fivebranes is possible only
because each is a magnetic source for the supergravity four-form field
strength $G=dC$~\cite{w:cy}. This provides an inflow mechanism to cancel the
anomaly on the lower dimensional space. In general, the magnetic
sources for $G$ are five-forms. Explicitly, if $0\leq x^{11}\leq\p\r$
parameterizes the orbifold interval, one has~\cite{nse}
\begin{equation}
   dG = J_1 \w \d(x^{11}) + J_2 \w \d(x^{11}-\p\r) 
          + \sum_i J_5^{(i)} \w \d(x^{11}-x^{(i)})
\label{Bid}
\end{equation}
where $J_1$ and $J_2$ are four-form sources on the two fixed planes
and $J_5^{(i)}$ is a delta-function four-form source localized at the
position of the $i$-th five-brane in $X$. The explicit one-form delta
functions give the positions of the orbifold fixed planes at
$x^{11}=0$ and $x^{11}=\p\r$ and the five-branes at $x^{11}=x^{(i)}$
in $\orb$.

Compactifying on $X\ts\orb$, we have the requirement that the net
charge in the internal space must vanish, since there is nowhere for
flux to escape. Equivalently, the integral of $dG$ over any five-cycle
in $X\ts\orb$ must be zero since $dG$ is exact. Integrating over the
orbifold interval then implies that the integral of $J_1+J_2+\sum_i
J_5^{(i)}$ over any four cycle in $X$ must vanish. Alternatively, this
means that the sum of these four-forms must be zero up to an exact
form, that is, they must vanish cohomologically. 

Explicitly, the source on each orbifold plane is proportional to 
\begin{equation}
   J_{n} \sim \tr F_n\w F_n - \hf \tr R\w R
\end{equation}
where $F_n$ for $n=1,2$ is the $E_8$ field strength on the $n$-th
fixed plane, while $R$ is the spacetime curvature. The full cohomology
condition can then be written as 
\begin{equation}
   \l(TX) = w(V_1) + w(V_2) + [W]
\label{cohocond}
\end{equation}
with
\begin{equation}
\begin{aligned}
   w(V) &=  - \frac{1}{60\cdot 8\pi^2} \tr_{\bf 248} F \w F \\
   \l(TX) &= \frac{1}{2}p_1(TX) 
       = - \frac{1}{2\cdot 8\p^2} \tr_{\bf 6} R \w R
\end{aligned}
\end{equation}
where the right-hand sides of these expressions really represent
cohomology classes, rather than the forms themselves. The traces are
in the adjoint ${\bf 248}$ of $E_8$ and the vector representation of
$SO(6)$. $[W]$ represents the total cohomology class of the
five-branes, which we will discuss in a moment. Note that $\l$ is half
the first Pontrjagin class. It is, in fact, an integer class because we
are on a spin manifold. On a Calabi--Yau threefold it is equal to the
second Chern class $c_2(TX)$, where the tangent bundle $TX$ is viewed
as an $SU(3)$ bundle and the trace is in the fundamental
representation. Thus, the cohomology condition simplifies to 
\begin{equation}
   [W] = c_2(TX) - w(V_1) - w(V_2)
\label{cocond}
\end{equation}

What do we mean by the cohomology class $[W]$? We recall that we
associated four-form delta function sources to the five-branes in
$X$. The class $[W]$ is then the cohomology class of the sum of all
these sources. Recall that the five-branes wrap on holomorphic curves
within the Calabi--Yau threefold. The sum of the five-branes thus
represents an integer homology class in $H_2(X,\ZZ)$. In general, one
can then use Poincar\'e duality to associate an integral cohomology
class in 
$H^4(X,\ZZ)$ to the homology class of the fivebranes, or also a de Rham
class in
$\HdR^4(X,\RR)$.
 This is
the class $[W]$ which enters the cohomology condition, though we will
throughout use the same expression $[W]$ for the 
integral 
homology class in
$H_2(X,\ZZ)$, 
the integral cohomology class in $H^4(X,\ZZ)$,  and the de Rham
cohomology class in
$\HdR^4(X,\RR)$.

\subsection{Homology classes and effective curves}

Let us now turn to analyzing the cohomology condition~\eqref{cocond}
in more detail. One finds that the requirement that $[W]$ correspond
to the homology class of a set of supersymmetric fivebranes puts a
constraint on the allowed bundle classes~\cite{dlow1,dlow2}. 

Since the sources are all four-forms, equation~\eqref{cocond} is
clearly a relation between de~Rahm cohomology classes
$\HdR^4(X,\RR)$. However, in fact, the sources are more restricted than
this. In general, they are all in integral cohomology classes. By this
we mean that their integral over any four-cycle in the Calabi--Yau
threefold gives an integer. (As noted above, this is even true when we
no longer have a Calabi--Yau threefold but only a spin manifold, and
$c_2(TX)$ is replaced by $\hf p_1(X)$.) The class $[W]$ is integral
because it is Poincar\'e dual to an integer sum of fivebranes, an
element of $H_2(X,\ZZ)$. Note that there is a general notion of the
integer cohomology group $H^p(X,\ZZ)$ which, in general, includes
discrete torsion groups such as $\ZZ_2$. This 
maps naturally to
de~Rahm cohomology $\HdR^p(X,\RR)$. However, it is important to note
that the 
map 
is not injective. Torsion elements in $H^p(X,\ZZ)$
are lost. The integral classes to which we refer in this paper are to
be identified with the images of $H^p(X,\ZZ)$ in $\HdR^p(X,\RR)$.

In general, $[W]$ cannot be just any integral class. We have seen
that supersymmetry implies that fivebranes are wrapped on holomorphic
curves within $X$. Thus $[W]$ must correspond to the homology class of
holomorphic curves. Furthermore, $[W]$ must refer to some physical
collection of fivebranes. Included in $H_2(X,\ZZ)$ are negative
classes like $-[C]$ where $C$ is, for example, a holomorphic curve in
$X$. These have cohomology representatives which would correspond to
the ``absence'' of a five-brane, contributing a negative magnetic
charge to the Bianchi identity for $G$ and negative
stress-energy. Such states are physically not allowed. The condition
that $[W]$ describes physical, holomorphic fivebranes further
constrains $c_2(TX)$, $w(V_1)$ and $w(V_2)$ in the cohomology
condition~\eqref{cocond}.

In order to formalize these constraints, we need to introduce some
definitions. We will use the following terminology.
\begin{itemize}
\item
A \textbf{curve} is a holomorphic complex curve in the Calabi--Yau
manifold. A curve is \textbf{reducible} if it can be written as the
union of two curves.
\item
A \textbf{class} is a homology class in $H_2(X,\ZZ)$ (or the Poincar\'e
dual cohomology class in $H^4(X,\ZZ)$). In general, it may or may not
have a representative which is a holomorphic curve. If it does, then
a class is \textbf{irreducible} if it has an irreducible
representative. Note that there may be other curves in the class which
are reducible, but the class is irreducible if there is at least one
irreducible representative.
\item
A class which can be written as a sum of irreducible classes with
arbitrary integer coefficients is called \textbf{algebraic}. 
\item
A class is \textbf{effective} if it can be written as the sum of
irreducible classes with positive integer coefficients. 
\end{itemize}
Note that we will occasionally use analogous terminology to 
refer to surfaces (or divisors) in $X$. These are holomorphic complex
surfaces in the Calabi--Yau threefold, so 
they 
have four real dimensions,
and 
their classes
lie in 
$H_4(X,\ZZ)$.

Physically, the above
definitions
correspond to the following. A curve $W$ describes a collection
of
supersymmetric fivebranes wrapped on holomorphic two-cycles in the
Calabi--Yau space. A reducible curve is the union of two or more separate
five-branes. A  general class in $H_2(X,\ZZ)$ has representatives
which are a general collection of five-branes, perhaps supersymmetric,
perhaps not, and maybe including ``negative'' fivebranes of the form
mentioned above. An algebraic class, on the other hand, has
representatives which are a collection of only five-branes wrapped on
holomorphic curves and so supersymmetric, but again includes the
possibility of negative fivebranes. Finally, an effective class has
representatives which are collections of supersymmetric fivebranes but
exclude the possibility of non-physical negative fivebrane states. 

From these conditions, we see that the constraint on $[W]$ is that we
must choose the Calabi--Yau threefold and the gauge bundles $V_1$ and
$V_2$ such that 
\begin{equation}
   \text{$[W]$ must be effective}
\end{equation}
As it stands, it is not clear that $[W]=c_2(TX)-w(V_1)-w(V_2)$ is 
algebraic, let alone effective. However, supersymmetry implies that
both the tangent bundle and the gauge bundles are holomorphic. There
is then a useful theorem that the classes of holomorphic bundles are
algebraic\footnote{This is a familiar result for Chern classes
(see~\cite{griffharr}). For $E_8$, or other groups, it can be seen by
taking any matrix representation of the group and treating it as a
vector bundle, that is, by embedding $E_8$ in $GL(n,\CC)$. The second
Chern class of the vector bundle is then algebraic and is some integer
multiple $p$ of the class $w(V)$, where the factor is related to the
quadratic Casimir of the representation. We conclude that $w(V)$ is
\textit{rationally algebraic}: it is integral, and a further integral
multiple of it is algebraic.}, and so $[W]$ is in fact
necessarily algebraic. However, there remains the condition that $[W]$
must be effective which does indeed constrain the allowed gauge
bundles on a given Calabi--Yau threefold.

\subsection{The theory on the fivebranes, $\cN=1$ gauge theories and
the fivebrane moduli space}

While two of the fivebrane dimensions are wrapped on a curve within
the Calabi--Yau manifold, the remaining four dimensions span
uncompactified Minkowski space. The low-energy massless degrees of
freedom on a given fivebrane consequently fall into four-dimensional
$\cN=1$ multiplets. At a general point in moduli space there are a set
of complex moduli $(m_i,\bar{m}_i)$ describing how the fivebrane curve
can be deformed within the Calabi--Yau three-fold. These form a set of
chiral multiplets. In addition, there is a single real modulus
$x^{11}$ describing the position of the fivebrane in the orbifold
interval. This is paired under supersymmetry with an axion $a$ which
comes from the reduction of the self-dual three-form degree of
freedom, $h$, on the fivebrane to form a further chiral
multiplet. When the fivebrane is non-singular, that is, does not
intersect itself, touch another fivebrane, or pinch, at any point, the
remaining degrees of freedom are a set of $U(1)$ gauge multiplets,
where the gauge fields also arise from the reduction of the self-dual
three-form. The number of $U(1)$ fields is given by the genus $g$ of
the curve. In summary, generically, we have
\begin{equation}
\begin{aligned}
   \text{chiral multiplets:}& \quad (x^{11},a),\, (m_i,\bar{m}_i) \\
   \text{vector multiplets:}& \quad 
         \text{$g$ multiplets with $U(1)^g$ gauge group}
\end{aligned}
\end{equation}
for each distinct fivebrane.

When the fivebrane becomes singular, new degrees of freedom can
appear. These correspond to membranes stretched between parts of the
same fivebrane, or the fivebrane and other fivebranes, which shrink and
become massless when the fivebrane becomes singular. They may be new
chiral or vector multiplets. In the following, we will not generally
identify all the massless degrees of freedom at singular
configurations but, rather, concentrate on describing the degrees of
freedom on the smooth parts of the moduli space. 

In conclusion, we have seen that fixing the Calabi--Yau manifold and
gauge bundles, in general, fixes an element $[W]$ of $H_2(X,\ZZ)$
describing the homology class of the holomorphic curve in $X$ on which
the fivebranes are wrapped. In order to describe an actual set of
fivebranes, $[W]$ must be effective, which puts a constraint on the
choice of gauge bundles. In general, there are a great many different
arrangements of fivebranes in the same homology class. The fivebranes
could move about within the Calabi--Yau threefold and also in the
orbifold interval. In addition, there can be transitions where branes
split and join. The net effect is that there is, in general, a
complicated moduli space of five-branes parameterizing all the
different possible combinations. In the low-energy effective theory on
the fivebranes, the moduli space is described by a set of chiral
multiplets. In order to describe the structure of this moduli space,
it is clear that we need to analyze the moduli space of all the
holomorphic curves in the class $[W]$, including the possibility that
each fivebrane can move in $\orb$ and can have a different value of
the axionic scalar $a$.


\section{Elliptically fibered Calabi--Yau manifolds} 

The moduli spaces we will investigate in detail in this paper are
those for five-branes wrapped on smooth elliptically fibered
Calabi--Yau threefolds $X$. Consequently, in this section we will
briefly summarize the structure of $X$, then identify the generic
algebraic classes and finally understand the conditions for these
classes to be effective.

\subsection{Properties of elliptically fibered Calabi--Yau threefolds}

An elliptically fibered Calabi--Yau threefold $X$ consists of a base
$B$, which is a complex surface, and an analytic map
\begin{equation}
   \p : X \to B
\label{XtoB}
\end{equation}
with the property that for a generic point $b \in B$, the fiber
$E_{b}=\pi^{-1}(b)$ is an elliptic curve. That is, $E_{b}$ is a
Riemann surface of genus one with a particular point, the origin $p$,
identified. In particular, we will require that there exist a global
section, denoted $\sigma$, defined to be an analytic map 
\begin{equation}
   \s : B \to X
\label{section}
\end{equation}
that assigns to every point $b\in B$ the origin $\s(b)=p\in
E_{b}$. We will sometimes refer to this as the zero section. The
requirement that the elliptic fibration have a section is crucial for
duality to F~theory. However, one notes that from the M~theory point
of view it is not necessary. 

In order to be a Calabi--Yau threefold, the canonical bundle of $X$
must be trivial. From the adjunction formula, this implies that the
normal bundle to the section, $N_{B/X}$, which is a line bundle over
$B$ and tells us how the elliptic fiber twists as one moves around the
base, must be related to the canonical bundle of the base, $K_B$. In
fact,  
\begin{equation}
   N_{B/X} = K_B.
\label{cL}
\end{equation}
Further conditions appear if one requires that the Calabi--Yau
threefold be smooth. The canonical bundle $K_B$ is then constrained so
that the only possibilities for the base manifold are as
follows~\cite{vm2,grassi}:
\begin{itemize}
\item
for a smooth Calabi--Yau manifold the base $B$ can be a del Pezzo
($dP_r$), Hirzebruch ($F_r$) or Enriques surface, or a blow-up of a
Hirzebruch surface. 
\end{itemize}
These are the only possibilities we will consider. The structure of
these surfaces is discussed in detail in an appendix
to~\cite{dlow2}. In the following, we will adopt the notation used
there.  

It will be useful to recall that, in general, there is a set of
points in the base at which the fibration becomes singular. These form
a curve, the discriminant curve $\D$, which is in the homology class
$-12K_B$, as can be shown explicitly by considering the 
Weierstrass form of the fibration.

\subsection{Algebraic classes on $X$}
\label{algclasses}

Since $[W]$ is algebraic, we need to identify the set of algebraic
classes on our elliptically fibered Calabi--Yau manifold. This was
discussed in~\cite{dlow1,dlow2}, but here we will be more explicit. It
will be useful to identify these classes both in $H_2(X,\ZZ)$
and $H_4(X,\ZZ)$. In general, the full set of classes will depend on
the particular fibration in question. However, there is a generic set
of classes which are always present, independent of the fibration, and
this is what we will concentrate on.

Simply because we have an elliptic fibration, the fiber at any given
point is a holomorphic curve in $X$. Consequently, one algebraic class
in $H_2(X,\ZZ)$ which is always present is the class of the fiber,
which we will call $F$. The existence of a section means there is also a
holomorphic surface in $X$. Thus the class of the section, which we
will call $D$, defines an algebraic class in $H_4(X,\ZZ)$. 

Some additional algebraic classes may be inherited from the base
$B$. In general, $B$ has a set of algebraic classes in
$H_2(B,\ZZ)$. One useful fact is that for all the bases which lead to
smooth Calabi--Yau manifolds, one finds that every class in
$H_2(B,\ZZ)$ is algebraic. This follows from the Lefschetz
theorem~\cite{griffharr} which tells us that we can identify algebraic
classes on a surface $S$ with the image of integer classes in the
Dolbeault cohomology $H^{1,1}(S)$. One then has the following
picture. In general, the image of $H^2(S,\ZZ)$ is a lattice of points
in $H^2(S,\RR)$. Choosing a complex structure on $S$ corresponds to
fixing an $h^{1,1}$-dimensional subspace within $H^2(S,\RR)$
describing the space $H^{1,1}(S)$. Generically, no lattice points will
intersect the subspace and so there are no algebraic classes on
$S$. The exception is when $h^{2,0}=0$, which is the case for all the
possible bases $B$. Then the subspace is the whole space $H^2(S,\RR)$ 
so all classes in $H^2(S,\ZZ)$ are algebraic. 

\begin{figure}
   \centerline{\psfig{figure=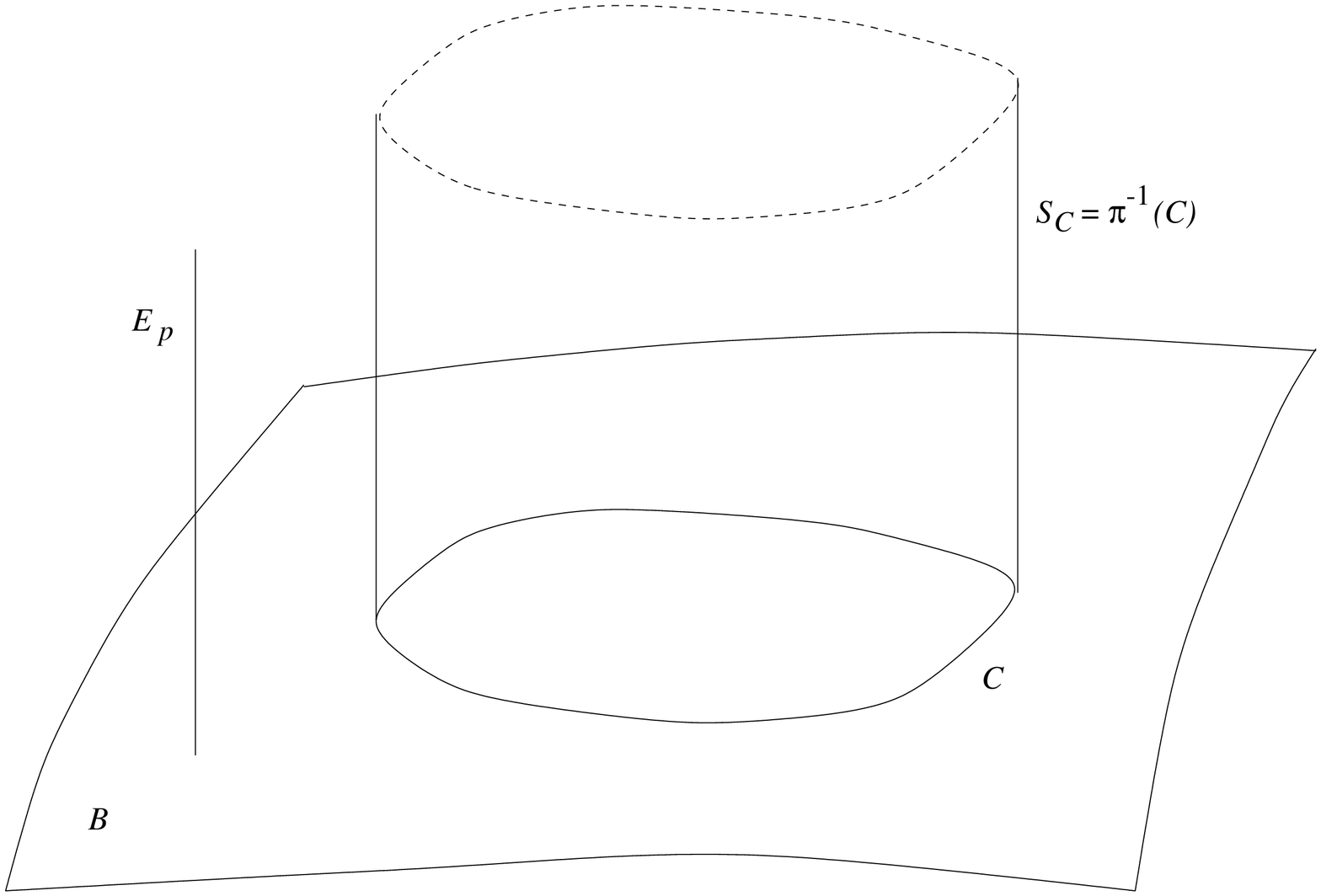,height=3in}}
   \caption{Generic algebraic classes on $X$, where $[C]=\s_*\O$ and
   the fiber class $[E_p]=F$ are in $H_2(X,\ZZ)$, and
   $[\p^{-1}(C)]=\p^*\O$ and base class $[B]=D$ are in $H_4(X,\ZZ)$.}
   \label{algclass}
\end{figure}

If $\O$ is an algebraic class in $H_2(B,\ZZ)$, there are two ways it
can lead to a class in $X$. First, one can use the section $\s$ to
form a class in $H_2(X,\ZZ)$. If $C$ is some representative of $\O$,
then the inclusion map $\s$ gives a curve $\s(C)$ in $X$. The homology
class of this curve in $H_2(X,\ZZ)$ is denoted by $\s_*\O$. Second, we
can use the projection map $\p$ to pull $\O$ back to a class in
$H_4(X,\ZZ)$. For a given representative $C$, one forms the fibered
surface $\p^{-1}(C)$ over $C$. The homology class of this surface in
$H_4(X,\ZZ)$ is then denoted by $\p^*\O$. This structure is indicated
in Figure~\ref{algclass}. 

In general, these maps may have kernels. For instance, two curves which
are non-homologous in $B$, might be homologous once one embeds them in
the full Calabi--Yau threefold. In fact, we will see that this is not
the case. One way to show this, which will be useful in the following,
is to calculate the intersection numbers between the classes in $H_2$
and $H_4$. We find
\begin{equation}
\begin{array}{rrccc}
   & & &\multicolumn{2}{c}{H_2(X,\ZZ)} \\
   & & \vline & \s_*\O' &  F  \\
   \cline{2-5}
   H_4(X,\ZZ) &
      \begin{array}{c} \pi^*\O \\ D \end{array} & 
      \vline &
      \begin{array}{c} \O\cdot \O' \\ K_B \cdot \O' \end{array} &
      \begin{array}{c} 0 \\ 1 \end{array}
\end{array}
\label{inters}
\end{equation}
where the entries in the first column are the intersections of classes
in $B$. The intersection of $\s_*\O'$ with $D$ is derived by adjunction,
recalling that the normal bundle to $B$ is $N_{B/X}=K_B^{-1}$. Two
classes are equivalent if they have the same intersection numbers. If
we take a set of classes $\O_i$ which from a basis of $H_2(B,\ZZ)$,
we see that the matrix of intersection numbers of the form given
in~\eqref{inters} is non-degenerate. Thus, for each nonzero $\O\in
H_2(B,\ZZ)$, we get nonzero classes $\s_*\O$ and $\p^*\O$ in
$H_2(X,\ZZ)$ and $H_4(X,\ZZ)$.  

As we mentioned, the algebraic classes we have identified so far are
generic, always present independently of the exact form of the
fibration. There are two obvious sources of additional
classes. Consider $H_4(X,\ZZ)$. First, we could have additional
sections non-homologous to the zero-section $\s$. Second, the
pull-backs of irreducible classes on $B$ could split so that
$\pi^*(\O)=\S_1+\S_2$. This splitting comes from the fact that there
can be curves on the base over which the elliptic curve degenerates,
for example, into a pair of spheres. New classes appear from wrapping
the four-cycle over either one sphere or the other. Now consider
$H_2(X,\ZZ)$. We see that the possibility of degeneration of the
fiber means that the fiber class $F$ can similarly split, with
representatives wrapped, for instance, on one sphere or the
other. Finally, the presence of new sections means there is a new way
to map curves from $B$ into $X$ and, in general, classes in
$H_2(B,\ZZ)$ will map under the new section to new classes in
$H_2(X,\ZZ)$.

In all our discussions in this paper, we will ignore these additional
classes. This will mean that our moduli space discussion is not in
general complete. However, this restriction will allow us to analyze
generic properties of the moduli space. In summary, we have identified
the generic algebraic classes $\O$ in $H_2(X,\ZZ)$ as classes in $B$
(since these are all algebraic for the bases in question) mapped via
the section into $X$, together with the fiber class $F$; while in
$H_4(X,\ZZ)$ the generic algebraic classes are the pull-backs
$\p^*(\O)$ of classes in $B$, together with the class $D$ of the
section. Furthermore, distinct algebraic classes in $B$ lead to
distinct algebraic classes in $X$.

\subsection{Effective classes on $X$}

We argued in the previous section that a generic algebraic class
$[W]$ in $H_2(X,\ZZ)$ on a general $X$ can be written as 
\begin{equation}
   [W] = \s_*\O + f F
\label{Wdecomp}
\end{equation}
where $\O$ is an algebraic class in $B$ (which is then mapped to $X$
via the section) and $F$ is the fiber class, while $f$ is some
integer. If $[W]$ is to be the class of a set of five-branes it must
be effective. What are the conditions, then, on $\O$ and $f$ such that
$[W]$ is effective?

We showed in~\cite{dlow2} that the following is true. First, for a
base which is any del Pezzo or Enriques surface, $[W]$ is effective if
and only if $\O$ is an effective class in $B$ and $f\geq 0$. Second,
this is also true for a Hirzebruch surface $F_r$, with the exception
of when $\O$ happens to contain the negative section $\cS$ and
$r\geq3$. Here, following the notation of~\cite{dlow2}, we write a
basis of algebraic classes on $F_r$ as the negative section $\cS$ and
the fiber $\cE$. In this case, there is a single additional
irreducible class $\s_*\cS-(r-2)\s_*\cE$. In this paper, for
simplicity, we will not consider these exceptional cases for which the
statement is untrue. Thus, under this restriction, we have that 
\begin{equation}
   W = \s_*\O + f F \text{ is effective in } X 
      \Longleftrightarrow \O \text{ is effective in } B \text{ and } f\geq 0
\label{thrm}
\end{equation}
This reduces the question of finding the effective curves in $X$ to 
knowing the generating set of effective curves in the base $B$. For
the set of base surfaces $B$ we are considering, finding such
generators is always possible (see for instance~\cite{dlow2}). 

The derivation of this result goes as follows. Clearly, if $\O$ is
effective in $B$ and $f$ is non-negative then since effective curves
in $B$ must map under the section to effective curves in $X$, we can
conclude that $[W]$ is effective. One can also prove that the converse
is true in almost all cases. One sees this as follows. First, unless a
curve is purely in the fiber, in which case $\O=0$, the fact that $X$
is elliptically fibered means that all curves $W$ project to curves in
the base. The class $[W]$ similarly projects to the class $\O$. The
projection of an effective class must be effective, thus if $[W]$ is
effective in $X$ then so is $\O$ in $B$. The only question then is
whether there are effective, irreducible, curves in $X$ with negative
$f$. To address this, we use the fact that any irreducible class in
$H_2(X,\ZZ)$ must have non-negative intersection with any effective
class in $H_4(X,\ZZ)$ unless all the representative curves are
contained within the representative surfaces. We start by noting that
if $\O$ is an effective class in $B$ then $\p^*\O$ must be an
effective class in $H_4(X,\ZZ)$. This can be seen by considering any
given representative of $B$ and its inverse image in $X$. From the
intersection numbers given in~\eqref{inters} and the generic form of
$[W]$~\eqref{Wdecomp} we see that, if $\O'$ is an effective class in
$B$, then  
\begin{equation}
\begin{aligned}
   \pi^*\O' \cdot [W] &= \O' \cdot \O \\
   D \cdot [W] &= K_B \cdot \O + f
\end{aligned}
\label{effinters}
\end{equation}
From the first intersection one simply deduces again that if $[W]$ is
effective then so is $\O$. Now suppose that $f$ is non-zero. Then
$W$ cannot be contained within $B$ and so, from the second expression,
we have $f\geq -K_B\cdot \O$. We recall that for del Pezzo and
Enriques surfaces $-K_B$ is nef, so that its intersection with any
effective class $\O$ is non-negative. Thus, we must have $f\geq 0$ for
$[W]$ to be effective. The exception is a Hirzebruch surface $F_r$ for
$r\geq 3$. We then have $-K_B\cdot \cE=2>0$ but $-K_B\cdot
\cS=2-r<0$. This allows the existence of effective classes of the
form $\s_*\cS+fF$ with $f$ negative. Indeed, the existence of such a
class can be seen as follows. Consider a representative curve $C$ of
$\cS$ in $F_r$ with $r\geq 3$ (in fact, the representative is
unique). It is easy to see that $C$ is topologically $\PP^1$ (see
equation~\eqref{genuscS} below). The surface $\p^{-1}(C)$ above $C$
should thus be an elliptic fibration over $\PP^1$. However, as shown in
equation~\eqref{pcS} below, in fact, $C$ is contained within the
discriminant curve $\D$ of the Calabi--Yau fibration. Thus all the
fibers over $C$ are singular. The generic singular fiber is a $\PP^1$,
suggesting that $S_C$ is a $\PP^1$ fibration over $\PP^1$. In fact
it can be shown that $S_C$ is indeed itself the Hirzebruch surface
$F_{r-2}$ (or a blow-up of such a surface). What class is our original
curve $C$ in the new surface $F_{r-2}$? If we write the classes of
$F_{r-2}$ as $\cS'$ and $\cE'$, we identify $\cE'=F$ since this is
just the fiber class of the $F_{r-2}$. In addition, one can show that
$\cS=\cS'+(r-2)\cE'$. However, we know that $\cS'$ itself is an
irreducible class, so $\cS'=\cS+(2-r)F$ is irreducible in
$H_2(X,\ZZ)$. Thus we see there is one new irreducible class with
negative $f$ which saturates the condition that $f\geq -(r-2)$.


\section{The moduli space for fivebranes wrapping the elliptic fiber
and the role of the axion} 

Probably the simplest example of a fivebrane moduli space is the case
where the fivebranes wrap only the elliptic fiber of the Calabi--Yau
threefold. By way of introduction to calculating moduli spaces, in
this section, we will consider this case, first for a single fivebrane
and then for a collection of fivebranes. These configurations are well
understood in the dual F-theory picture as collections of
D3-branes~\cite{fmw}. We end the section with a discussion of the connection
between our results and the F-theory description. 

\subsection{$[W]=F$}

If it wraps a fiber only once, the class of the fivebrane curve is
simply given by 
\begin{equation}
   [W] = F
\end{equation}
A fivebrane wrapping any of the elliptic fibers will be in this
class. One might imagine that there are other fivebranes in this class,
where not all the fivebrane lies at the same point in the Calabi--Yau
threefold. Instead, as one moves along the fivebrane in the fiber
direction, the fivebrane could have a component in the base
directions. However, if the curve is to be holomorphic, every point in
the fivebrane curve must lie over the same point in the
base. Similarly, in order to preserve $\cN=1$ supersymmetry, the
brane must be parallel to the orbifold fixed planes, so it is also at
a fixed point in the orbifold. Since these position moduli are
independent, the moduli space appears to be $B\ts\orb$. The two
complex coordinates on $B$ form a pair of chiral $\cN=1$
superfields. The metric on this part of the moduli space should simply
come from the K\"ahler metric on the base $B$.  

However, we have, thus far, ignored the axionic scalar, $a$, on the
fivebrane world volume. We have argued that this is in a chiral
multiplet with the orbifold modulus $x^{11}$. Furthermore, it is
compact, describing an $S^1$. However, at the edges of the orbifold
this changes. It has been argued in~\cite{gh,sw} that there is a
transition when a fivebrane reaches the boundary. At the boundary, the
brane can be described by a point-like $E_8$ instanton. New low-energy
fields then appear corresponding to moving in the instanton moduli
space. Similarly, some of the fivebrane moduli disappear. Throughout
this transition the low-energy theory remains $\cN=1$. Thus, since the
$x^{11}$ degree of  freedom disappears in the transition, so must the
axionic degree of freedom. Consequently the axionic $S^1$ moduli space
must collapse to a point at the boundary. We see that the full
$(x^{11},a)$ moduli space is just the fibration of $S^1$ over the
interval $\orb$, where the $S^1$ is singular at the boundaries, that
is, the orbifold and axion part of the moduli space is simply
$S^2=\PP^1$.

The fact that the axionic degree of freedom disappears on the
boundary can be seen in another way. In the fivebrane equation of
motion, one can write the self-dual three-form field strength $h$ in terms
of a two-form potential, $b$, in combination with the pull-back onto the
fivebrane worldvolume of the eleven-dimensional three-form potential
$C$ as~\cite{fbraneeom} 
\begin{equation}
   h = db - C
\label{hdef}
\end{equation}
Under the $Z_2$ orbifold symmetry $C$ is odd unless it has a component
in the direction of the orbifold. Since the fivebrane must be parallel
to the orbifold fixed-planes this is not the case. This implies that $h$
must also be odd. Consequently, $h$ must be zero on the orbifold fixed
planes implying that the axion $a$ also disappears on the boundary. 

In summary, the full moduli space is given locally by 
\begin{equation}
   \cM(F) = B \ts \PPa. 
\label{Fmod}
\end{equation}
where the subscript on $\PPa$ denotes that this part of moduli space
describes the axion multiplet. Globally, this $\PP^1$ could twist as we
move in $B$; so $\cM(F)$ is really a $\PP^1$ bundle over $B$. We will
return to this point below. What about the vector degrees of freedom?
Since the fiber is elliptic, the fivebrane curve must be topologically
a torus. Thus we have 
\begin{equation}
    g=\genus{W}=1
\end{equation}
and there is a single $U(1)$ vector multiplet in the low-energy
theory.

\subsection{$[W]=nF$}
\label{sec:fF}

The generalization to the case where the fivebrane class is a number of
elliptic fibers is straightforward. The class 
\begin{equation}
   [W] = fF
\end{equation}
where $f\geq 1$, means we have a collection of curves which wrap the
fiber $f$ times. In general, we could have one component which wraps
$f$ times or two or more components each wrapping a fewer number of
times. In the limiting case, there are $f$ distinct components each
wrapping only once. A single component must wrap all at the same
point in the base. In addition, it must be at a fixed point in the
orbifold interval and must have a single value of the axionic
scalar. Two or more distinct components can wrap at different points
in the base and have different values of $x^{11}$ and $a$. 

As homology cycles, there is no distinction between the case where
some number $n$ of singly wrapped components overlap, lying at the
same point in the base, and the case where there is a single component
wrapping $n$ times. Both cases represent the same two-cycle in the
Calabi--Yau manifold. Physically, they could be distinguished if the
$n$ singly wrapped components were at different points in $\orb$ or
had different values of the axion. However, if the values of $x^{11}$
and $a$ were also the same, by analogy with D~branes, we would expect
that we could not then distinguish, in terms of the scalar fields on
the branes, the $n$ singly wrapped fivebranes from a single brane
wrapped $n$ times. From the discussion in the last section, each
singly wrapped fivebrane has a moduli space given locally by
$B\times\PPa$. Thus for $f$ components, we expect the full scalar
field moduli space locally has the form
\begin{equation}
   \cM(fF) = \left(B\times\PPa\right)^f/\ZZ_f
\label{fFmod}
\end{equation}
where we have divided out by permutations since the fivebranes are
indistinguishable. The ambiguous points in moduli space, which could
correspond to a number of singly wrapped fivebranes or a single
multiply wrapped fivebrane, are then the places where two or more of
the points in the $f$ $B\times\PPa$ factors coincide. Note, in
addition, that this is again only the local structure of $\cM(fF)$. We
do not know how the $\PPa$ factors twist as we move the fivebranes in
the base. Thus, globally, $\cM(fF)$ is the quotient of a $(\PP^1)^f$
bundle over $B^f$.

In a similar way, the gauge symmetry on the fivebranes also follows by
analogy with D-branes. At a general point in the moduli space, we have
$n$ distinct fivebranes each wrapping a torus and so, as in the
previous section, each with a single $U(1)$ gauge field. When two
branes collide in the Calabi--Yau threefold, and are at the same point
in the orbifold and have the same value of the axion, we expect the
symmetry enhances to $U(2)$. The new massless states come from
membranes stretched between the fivebranes. The maximal enhancement is
when all the fivebranes collide and the group becomes $U(n)$.

\subsection{Duality to F~theory and twisting the axion}
\label{axiontwist}

The results of the last two sections are extremely natural from the
F~theory point of view. It has been argued that fivebranes wrapping an
elliptic fiber of $X$ correspond to threebranes spanning the flat
$M_4$ space on the type IIB side~\cite{fmw}. To understand the
correspondence, we first very briefly review the relation between M
and F~theory~\cite{vafa,vm1,vm2}. 

The duality states that the heterotic string on an elliptically
fibered Calabi--Yau threefold $X$ is dual to F~theory on a Calabi-Yau
fourfold $X'$ fibered by K3 over the same base $B$. The M~theory limit
of the heterotic string we consider here is consequently also dual to
the same F~theory configuration. In addition, the duality requires
that the K3 fibers should themselves be elliptically fibered. This
means that the fourfold $X'$ also has a description as an elliptic
fibration over a threefold base $B'$. Since the base of an
elliptically fibered K3 manifold is simply $\PP^1$, this implies that
$B'$ must be a $\PP^1$ fibration over $B$. As a type IIB background,
the spacetime is $B'\times M_4$, where $M_4$ is flat Minkowski
space. The complex structure of the elliptic fibers of $X'$ then
encode how the IIB scalar doublet, the dilaton and the Ramond-Ramond
scalar, vary as one moves over the ten-dimensional manifold $B'\times
M_4$. As such, they describe some configuration of seven-branes in
type IIB. 

M~theory fivebranes which wrap the elliptic fiber in $X$, map 
to threebranes spanning $M_4$ in the dual F~theory vacuum. As such,
the three-brane is free to move in the remaining six compact
dimensions. Thus we expect that the threebrane moduli space is simply
$B'$. However, we have noted that $B'$ is a $\PP^1$ fibration over
$B$. Thus we see that locally the moduli space as calculated on the
F~theory side exactly coincides with the moduli space of the fivebrane
given in~\eqref{Fmod} above. The $\PP^1$ fiber in $B'$ is precisely
the orbifold coordinate $x^{11}$ together with the axion $a$. For a
collection of $f$ threebranes, we expect the moduli space is simply
promoted to the symmetric product ${B'}^f/Z_f$. Again, locally, this
agrees with the moduli space~\eqref{fFmod} of the corresponding M theory
fivebranes. Similarly, it is well known that a threebrane carries a
single $U(1)$ gauge field, as does the M~theory fivebrane. For a
collection of $f$ threebranes this is promoted to $U(f)$, which was
really the motivation for our claim for the vector multiplet
structure calculated in the M~theory picture. 

In general, the arguments given in the previous two sections were only
sufficient to give the local structure of the axion multiplet part of
the fivebrane moduli space. We did not determine how the axion fiber
$\PPa$ twisted as one moved the fivebrane in the Calabi--Yau
manifold. From duality with F~theory, we have seen that, in general,
we expect this twisting is non-trivial. In fact, it can also be
calculated from the M~theory side. We will not give the details here
but simply comment on the mechanism. A full description will be given
elsewhere~\cite{atwist}. The key is to recall that the self-dual
three-form on the fivebrane~\eqref{hdef} depends on the pull-back of the
supergravity three-form potential $C$. This leads to holonomy for the
axion degree of freedom as one moves the fivebrane within the
Calabi--Yau threefold. The holonomy can be non-trivial if the field
strength $G$ is non-trivial. However, from the modified Bianchi
identity~\eqref{Bid}, we see that this is precisely the case when
there are non-zero sources from the boundaries of $\orb$ and also from
the fivebranes in the bulk. In general, one can calculate how the axion
twists, and hence how $\PPa$ twists, in terms of the different sources.

This phenomenon is interesting but not central to the structure of the
fivebrane moduli spaces, such as the dimension of the space, how its
different branches intersect, or waht is the the genus of the
fivebrane curve. Thus, for simplicity, in the rest of this paper we
will ignore the issue of how $a$ twists as one moves a given
collection of fivebranes within the Calabi--Yau
manifold. Consequently, the moduli spaces we quote will strictly only
be locally correct for the axion degrees of freedom. So that it is
clear where the extra global structure can appear, we will always
label the $\PP^1$ degrees of freedom associated with the axions as
$\PPa$.


\section{Two examples with fivebranes wrapping curves in the base}

The discussion of the moduli space becomes somewhat more complicated
once one includes classes where the fivebrane wraps a curve in the
base  manifold. Again, we will take two simple examples to illustrate
the  type of analysis one uses. In both cases, we will assume, for
specificity, that the base manifold is a $dP_8$ surface, though the
methods of our analysis would apply to any base $B$. Throughout, we
will use the notation and results of~\cite{dlow2}. A $dP_8$ surface is
a $\PP^2$ surface blown up at eight points, $p_1,\dots,p_8$. In
general there are nine algebraic classes in the base: the class $l$
inherited from the class of lines in the $\PP^2$ and the eight classes
of the blown-up points $E_1,\dots,E_8$. In the following, we will
often describe curves in $dP_8$ in terms of the corresponding plane
curve in $\PP^2$. 

\subsection{$[W]=\s_*l-\s_*E_1$}
\label{exbase1}

We first take an example where the fivebrane class includes no fiber 
components
\begin{equation}
   [W] = \s_*l - \s_*E_1
\end{equation}
where $\s_*l$ and $\s_*E_1$ are the images in the Calabi--Yau manifold
of the corresponding classes in the base. Since $\O=l-E_1$ is an
effective class in the base, (see~\cite{dlow2}), from~\eqref{thrm} we
see that $[W]$ is effective in $X$, as required. 

\begin{figure}
   \centerline{\psfig{figure=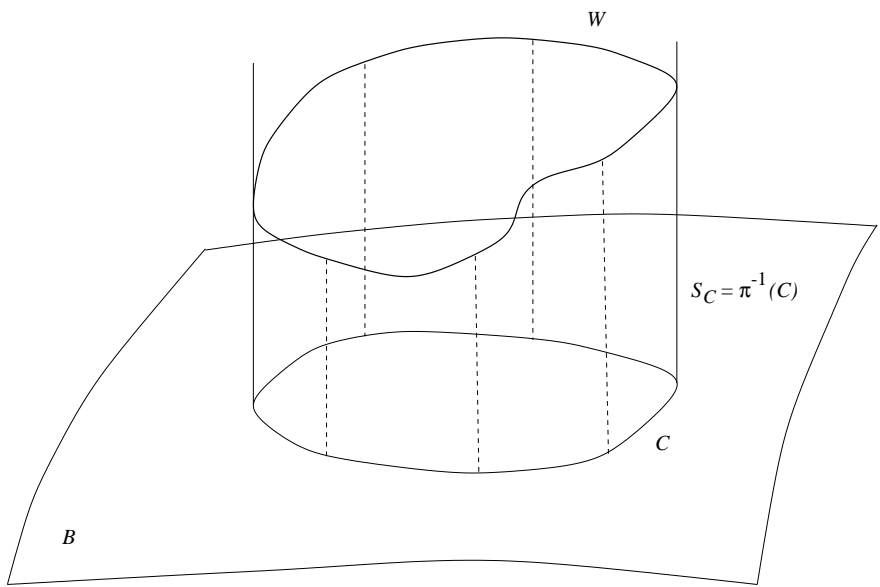,height=3in}}
   \caption{The curve $W$ and its image in the base $C$}
   \label{piC}
\end{figure}

If we knew that the curve lay only in the base, the moduli space would
then simply be the space of curves in a $dP_8$ surface in the class
$l-E_1$, which is relatively easy to calculate. In general, however,
$W$ lies somewhere in the full Calabi--Yau threefold. The fact that
its homology class is the image of a homology class in the base does
not imply that $W$ is stuck in $B$. Nonetheless, we do know that, under
the projection map $\p$ from $X$ to $B$, the curve $W$ must project
onto a curve $C$ in the base as shown in Figure~\ref{piC}. Furthermore
the class of $C$ must be $\O=l-E_1$ in $B$. What we can do is find the
moduli space of such curves $C$ in the base and then ask, for each
such $C$, what set of curves $W$ in the full Calabi--Yau manifold would
project onto $C$. That is to say, the full moduli space should have a
fibered structure. The base of this space will be the moduli space of
curves $C$ in $B$, while the fiber above a given curve is the class of
$W$ in $X$ which projects onto the given $C$. 

In our example, the moduli space of curves $C$ in the class $\O=l-E_1$ 
is relatively easy to analyze. In $\PP^2$, $\O$ describes the class of
lines through one point, $p_1$. A generic line in $\PP^2$ is a
homogeneous polynomial of degree one,
\begin{equation} 
   ax + by + cz = 0
\end{equation}
where $[x,y,z]$ are homogeneous coordinates on $\PP^2$. Since the
overall coefficient is irrelevant, a given line is fixed by giving
$[a,b,c]$ up to an overall scaling. Thus the moduli space of lines is
itself $\PP^2$. Furthermore, we see that a given point in the line is
specified by fixing, for instance, $x$ and $y$ up to an overall
scaling. Consequently, we see that, topologically, a line in $\PP^2$ is
just a sphere $\PP^1$. For the class $\O$, we further require that the
line pass through a given point $p_1=[x_1,y_1,z_1]$. This provides a
single linear constraint on $a$, $b$ and $c$, 
\begin{equation} 
   ax_1 + by_1 + cz_1 = 0
\end{equation}
We now have only the set of lines radiating from $p_1$ and the moduli
space is reduced to $\PP^1$. Topologically, the line in $\PP^2$ is
still just a sphere and, generically, its image in $dP_8$ will also
be a sphere.  

\begin{figure}
   \centerline{\psfig{figure=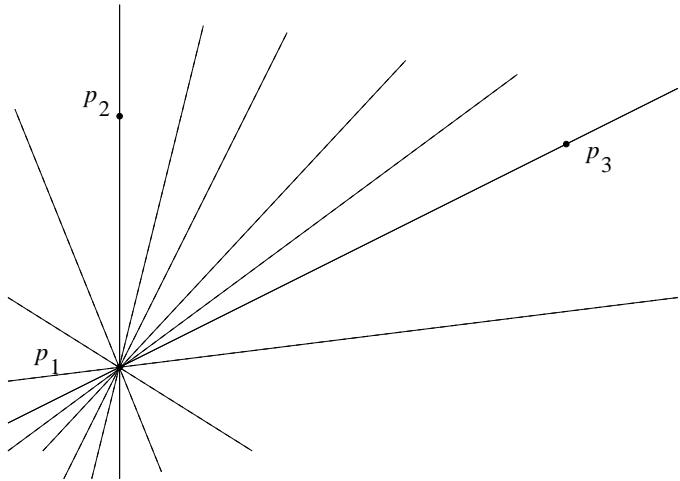,height=2.5in}}
   \caption{The moduli space of lines in the class $l-E_1$. The solid
   lines represent two examples of the special case where the proper
   transform of the line splits into two components}
   \label{lE1}
\end{figure}

\begin{figure}
   \centerline{\psfig{figure=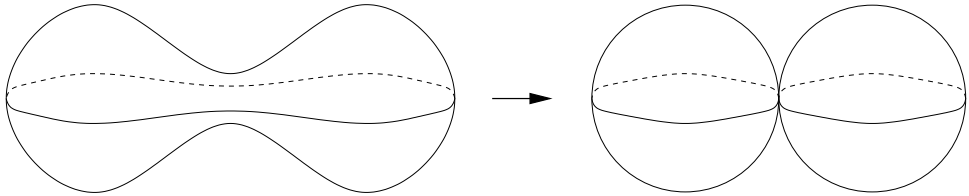,height=1in}}
   \caption{Splitting a single sphere into a pair of spheres}
   \label{Cpinch}
\end{figure}

There are, however, seven special points in the moduli space. A
general line passing through $p_1$ will not intersect any other blown-up
point. However, there are seven special lines radiating from $p_1$
which also pass through a second blown-up point. (To be a $dP_8$
manifold, the eight blow-up points must be in general position, so no
three are ever in a line.) This is shown in Figure~\ref{lE1}. Let us
consider one of these seven lines, say the one which passes through
$p_2$. The transform of such a line to $dP_8$ splits into two curves 
\begin{equation}
   C = C_1 + C_2
\end{equation}
The first component $C_1$ projects back to the line in $\PP^2$. The
second component corresponds to a curve wrapping the blown up $\PP^1$ at
$p_2$ and so has no 
analog 
in $\PP^2$. Specifically, the classes of
the two curves are
\begin{equation}
   [C_1] = l - E_1 - E_2,  \qquad
   [C_2] = E_2 
\label{Csplit}
\end{equation} 
Using the results in the Appendix to~\cite{dlow2}, we see that
\begin{equation}
   [C_1] \cdot [C_1] = [C_2] \cdot [C_2] = -1
\end{equation}
It follows that both curves are in exceptional classes in $dP_8$ and
so cannot be deformed within the base. Hence, no new moduli for moving
in the base appear when the curve splits. 
From the form of~\eqref{Csplit}, we see that, when the curve splits,
$C_1$ remains a line in $\PP^2$ so is topologically still a sphere,
while $C_2$ wraps the blown up $\PP^1$ and, so, is also topologically a
sphere. Furthermore, the intersection number 
\begin{equation}
   [C_1]\cdot [C_2]=1
\end{equation}
implies that the two spheres intersect at one point. What has
happened is that the single sphere $C$ has pinched off into a pair of
spheres as shown in Figure~\ref{Cpinch}. In summary, for the moduli
space of curves $C$ in the base, in the homology class $\O=l-E_1$, we
have,
\begin{equation}
\begin{array}{c|cc}
   [C]    & \text{genus} & \text{moduli space} \\
   \hline
   l-E_1  
      & 0 & \PP^1 - 7\text{ pts.} \\
   (l-E_1-E_i) + (E_i) 
      & 0+0 & \text{single pt.}
\end{array}
\label{Cmodspace}
\end{equation}
where in the second line $i=2,\dots,8$. 

The next step is to find, for a given curve $C$, how many curves $W$
there are in the full Calabi--Yau space which project onto
$C$. Furthermore, $W$ must be in the homology class
$\s_*l-\s_*E_1$. Let us start with a curve $C$ at a generic point in the
moduli space~\eqref{Cmodspace}, that is, a point in the first line of the
table where the curve has not split. Any curve $W$ which projects onto
$C$ must lie somewhere in the space of the elliptic fibration over
$C$. Thus, we are interested in studying the complex surface
\begin{equation}
   S_C = \pi^{-1}(C)
\end{equation}
This structure is shown in Figure~\ref{piC}. By
definition, this surface is an elliptic fibration over $C$ which,
means it is a fibration over $\PP^1$. In general, the surface will
have some number of singular fibers. This is equal to the intersection
number between the discriminant curve $\D$, which gives the position
of all the singular fibers on $B$, and the base curve $C$. Recall that
$[\D]=-12K_B$. Using the results summarized in the Appendix
to~\cite{dlow2}, since the base is a $dP_8$ surface and intersection
numbers depend only on homology classes, we have 
\begin{equation}
   [\D] \cdot [C] 
      = 12\left(3l-E_1-\dots-E_8\right) \cdot \left(l-E_1\right)
      = 24
\end{equation}
Thus we see that, generically, $S_C$ is an elliptic fibration over
$\PP^1$ with 24 singular fibers. This implies 
~\cite{griffharr}
that
\begin{equation}
   S_C \text{ is a K3 surface}
\end{equation}

The curve $C$ is the zero section of the fibration. Further,
projection gives us a map from the actual curve $W$ to its image $C$
in the base. The projection only wraps $C$ once, so, since $C$ is not
singular, the map is invertible and $W$ must also be a section of
$S_C$. Our question then simplifies to asking: what is the moduli space
of sections of $S_C$ in the class $\s_*l-\s_*E_1$? 

To answer this question, we start by identifying the algebraic classes
in $S_C$. We know that we have at least two classes inherited from the
Calabi--Yau threefold: the class of the zero section $C$, which we
write as $D_C$, and the class of the elliptic fiber,
$F_C$. Specifically, under the inclusion map 
\begin{equation}
   i_C : S_C \to X
\end{equation}
$D_C$ and $F_C$ map into the corresponding classes in $X$
\begin{equation}
\begin{aligned}
   i_{C*}D_C &= [C] = \s_*l - \s_*E_1 \\
   i_{C*}F_C &= F
\end{aligned}
\label{K3inclmap}
\end{equation}
where $i_{C*}$ is the map between classes
\begin{equation}
   i_{C*} : H_2(S_C,\ZZ) \to H_2(X,\ZZ)
\end{equation}
These are the only relevant generic classes in $X$. However, there may
be additional classes on $S_C$ which map to the same class in
$X$ so that the map $i_{C*}$ has a kernel. That is to say, two curves
which are homologous in $X$ may not be homologous in $S_C$. However,
we note that a generic K3 surface would have no algebraic classes
since $h^{2,0}\neq 0$ (see the discussion in
section~\ref{algclasses}). Given that in our case of an elliptically
fibered K3 with section we have at least two algebraic classes, the
choice of complex structure on $S_C$ cannot be completely
general. However, generically, we have no reason to believe that there
are any further algebraic classes. For particular choices of complex
structure additional classes may appear but, since here we are
considering the generic properties of the moduli space, we will ignore
this possibility. 

Now, we require that $W$, like $C$, is also in the class
$\s_*l-\s_*E_1$ in the full Calabi--Yau space. This immediately implies,
given the map~\eqref{K3inclmap}, that $W$ is also in the class $D_C$
of the zero section $C$ in $S_C$. Furthermore, we can calculate the
self-intersection number of this class within $S_C$. This can be done
as follows. Recall that the Riemann--Hurwitz formula~\cite{griffharr}
applied to the curve $C$ states that 
\begin{equation}
   2g - 2 = \deg K_C 
\label{RHformula}
\end{equation}
where $g$ is the genus and $K_C$ is cohomology class of the canonical
bundle of $C$. The adjunction formula~\cite{griffharr} then gives 
\begin{equation}
   \deg K_C = \left(K_{S_C} + D_C \right) \cdot D_C
\label{adj}
\end{equation}
where $K_{S_C}$ is the canonical class of the K3 surface $S_C$. Using
the fact that the canonical class of a K3 surface is zero,
$K_{S_C}=0$, and that $C$ is a sphere so $g=0$, it follows
from~\eqref{RHformula} and~\eqref{adj} that 
\begin{equation}
   D_C \cdot D_C = - 2
\label{Omegasq1}
\end{equation}
This implies that the section cannot be deformed at all within the
surface $S_C$. In conclusion, we see that there is, generically, no
moduli space of curves $W$ which project onto $C$. Rather, the only
curve in $S_C$ in the class $\s_*l-\s_*E_1$ is the section $C$
itself. We see that, generically, the curve $W$ can only move in the
base of $X$ and cannot be deformed in a fiber direction. 

Recall that a fivebrane wrapped on $W$ also has a modulus describing
its position in $\orb$, as well as the axionic modulus. Together, as
was discussed in the previous section, these form a $\PP^1$ moduli
space. Thus, we conclude that the moduli space associated with a
generic curve in~\eqref{Cmodspace} is locally simply 
\begin{equation}
   \cM_{\text{generic}} = \left(\PP^1 - \text{7 pts.}\right) \ts \PPa
\label{genericmod}
\end{equation}
As discussed above, since the axion can be twisted, globally, this
extends to a $\PPa$ bundle over $\PP^1$. Physically, we have a
single fivebrane wrapping an irreducible curve in the Calabi--Yau
threefold, which lies entirely within the base $B$. The curve can be
deformed in the base, which gives the first factor in the moduli
space, but cannot be deformed in the fiber direction. It can also move
in the orbifold interval and have different values for the axionic
modulus, which gives the second factor in~\eqref{genericmod}. Since
the curve has genus zero, there are no vector fields in the low-energy
theory. 

Thus far we have discussed the generic part of the moduli space. The
full moduli space will have the form
\begin{equation}
   \cM(\s_*l-\s_*E_1) = \left(\PP^1 - \text{7 pts.}\right) \ts \PPa
        \cup \cM_{\text{non-generic}}
\end{equation}
where the additional piece $\cM_{\text{non-generic}}$ describes the
moduli space at each of the special 7 points where the curve $C$
splits into two components. To analyze this part of the moduli space,
we must consider each component separately, but we can use the same
procedure we used above. The fact that the image $C$ splits, means
that the original curve $W$ must also split in $X$
\begin{equation}
   W = W_1 + W_2
\label{simplesplit}
\end{equation}
with $C_1$ being the projection onto the base of $W_1$ and $C_2$ the
projection of $W_2$. Let us consider the case where the line in $\PP^2$ also
intersects $p_2$. Then the homology classes of $W_1$ and $W_2$ must
split as 
\begin{equation}
   [W_1] = \s_*l - \s_*E_1 - \s_*E_2  \qquad
   [W_2] = \s_*E_2 
\label{Wsplit}
\end{equation} 
Note that one might imagine adding $nF$ to $[W_1]$ and $-nF$ to
$[W_2]$, still leaving the total $[W]$ unchanged and having the
correct projection onto the base. However, from~\eqref{thrm}, we see
that one class would not then be effective and so, since $W_1$ and
$W_2$ must each correspond to a physical fivebrane, such a splitting
is not allowed.

If we start with $W_1$, to find the curves in $X$ which project onto
$C_1$ and are in the homology class $\s_*l-\s_*E_1-\s_*E_2$, we begin,
as above, with the surface $S_{C_1}=\p^{-1}(C_1)$ above
$C_1$. Calculating the number of singular fibers, we find
\begin{equation}
   \D \cdot [C_1] 
       = 12\left(3l-E_1-\dots-E_8\right) \cdot \left(l-E_1-E_2\right)
       = 12
\end{equation}
Since $C_1$ is a sphere, we have an elliptic fibration over $\PP^1$
with 12 singular fibers, which implies that~\cite{griffharr}
\begin{equation}
   S_{C_1} \text{ is a $dP_9$ surface}
\end{equation}
Similarly, if we consider $[C_2]=E_2$, 
\begin{equation}
   \D \cdot [C_2] 
       = 12\left(3l-E_1-\dots-E_8\right) \cdot E_2
       = 12
\end{equation}
Since the curve $C_2$ is also a sphere, it follows that we again have
an elliptic fibration over $\PP^1$ with 12 singular fibers, and hence
we also have
\begin{equation}
   S_{C_2} \text{ is a $dP_9$ surface}
\end{equation}
Thus, we are considering the degeneration of the K3 surface $S_C$,
which had 24 singular fibers, into a pair of $dP_9$ surfaces $S_{C_1}$
and $S_{C_2}$, each with 12 singular fibers. 
On a given $dP_9$ surface, say $S_{C_1}$, we are
guaranteed, as for the K3 surface, that there are at least two
algebraic classes, the section class $D_{C_1}$ and the fiber class
$F_{C_1}$. However, the $dP_9$ case is more interesting than the case
of a K3 surface since there are always other additional algebraic
classes. On a $dP_9$ surface, $h^{2,0}=0$. Consequently, as was
discussed in section~\ref{algclasses}, whatever complex structure one
chooses, all classes in $H_2(dP_9,\ZZ)$ are algebraic. Thus, one finds
that the algebraic classes on $dP_9$ form a 10-dimensional
lattice. Since there are only two distinguished classes on the
Calabi--Yau threefold (namely $\s_*l-\s_*E_1$ and the fiber class
$F$), this implies that distinct classes in $S_{C_1}$ must map to the
same class in $X$. That is to say, curves which are not homologous in
$S_{C_1}$ are homologous once one considers the full threefold $X$. 

The full analysis of the extra classes on $S_{C_1}$ will be considered
in section~\ref{sec:dP9}. In our particular case, it will turn out
that, for $W_1$ to be in the same class as $C_1$ in the full
Calabi--Yau threefold, it must also be in the same class within
$S_{C_1}$. Thus, we are again interested in the moduli space of the
section class $D_{C_1}$ in $S_{C_1}$. Now, we recall (see, for
instance, the Appendix to~\cite{dlow2}) that the canonical class for a
$dP_9$ is simply $K_{S_{C_1}}=-F_D$. So by the analogous calculation
to~\eqref{Omegasq1}, using the fact that $D_{C_1}\cdot F_D=1$ since
$C_1$ is a section, we have that 
\begin{equation}
   D_{C_1} \cdot D_{C_1} = -1
\label{dP9sec}
\end{equation}
This means that the curve $C_1$ cannot be deformed within
$S_{C_1}$. Thus, as in the K3 case, the only possible $W_1$ is the
section $C_1$ itself. 

An identical calculation goes through for the other component
$W_2$. Furthermore, the analysis is the same at each of the other six
exceptional points in moduli space. Given that the curve has split
into two components at each of these points, we have two separate moduli
describing the position of each component in $\orb$ as well as two
moduli describing the axionic degree of freedom for each
component. It follows that 
\begin{equation}
   \cM_{\text{non-generic}} = 7\left(\PPa \ts \PPa\right)
\label{nongeneric}
\end{equation}
where $7(\PP^1\ts\PP^1)=\PP^1\ts\PP^1\cup\dots\cup\PP^1\ts\PP^1$. We
find, then, that the full moduli space has a branched structure, 
\begin{equation}
   \cM(\s_*l-\s_*E_1) = \left(\PP^1 - \text{7 pts.}\right) \ts \PPa
        \cup 7\left(\PPa \ts \PPa\right)
\label{lE1mod}
\end{equation}
where, globally, the first component, $\cM_{\text{generic}}$, in
fact, extends to a $\PP^1$ bundle over $\PP^1$. We can also describe
the way each copy of $\PPa\ts\PPa$ is attached to
$\cM_{\text{generic}}$: the diagonal of $\PPa\ts\PPa$, the set of
points where the two components intersect, is glued to a fiber of the
$\PPa$ bundle $\cM_{\text{generic}}$.

\begin{figure}
   \centerline{\psfig{figure=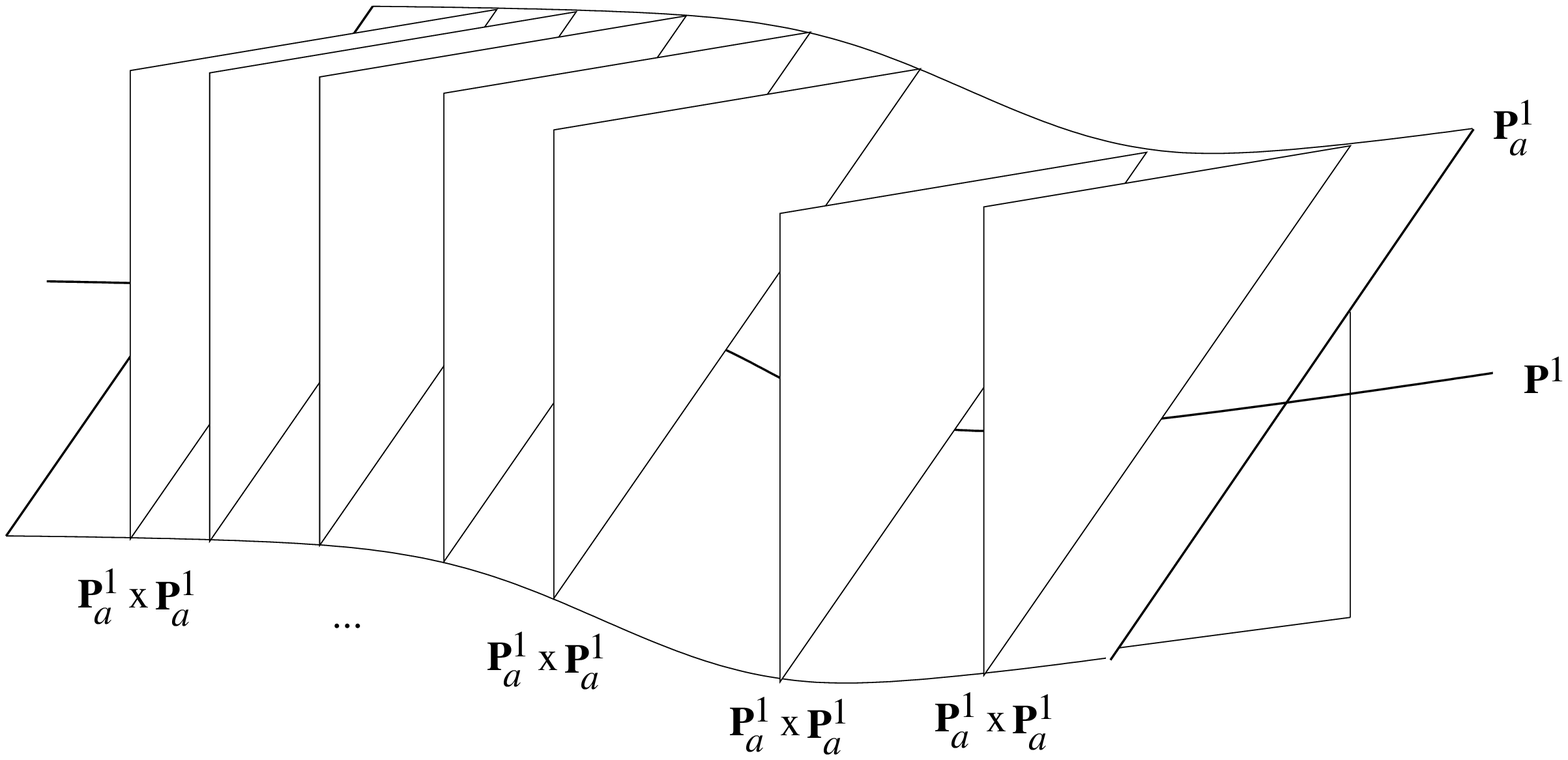,height=2.8in}}
   \caption{The moduli space $\cM(\s_*l-\s_*E_1)$.}
   \label{lE1modfig}
\end{figure}

Physically, as we discussed above, at a generic point in the moduli
space we have a single fivebrane wrapping a curve which lies solely in
the base of the Calabi--Yau threefold and is topologically a
sphere. The curve can be moved in the base and in $\orb$ but not in
the fiber direction. In moving around the base there are seven special
points where the fivebrane splits into two curves intersecting at one
point, as in Figure~\ref{Cpinch}. These are each fixed in both the
base and the fiber of the Calabi--Yau manifold, but can now each move
independently in $\orb$. The two fivebranes can then be separated so
that they no longer intersect. 
In making the transition from one of these
branches of the moduli space to the case where there is a single
fivebrane, the two fivebranes must be at the same point in $\orb$ and
have the same value of the axionic scalar $a$. They can then combine
and be deformed away within the base as a single curve. This structure
is shown in Figure~\ref{lE1modfig}. Note that, unlike the pure fiber
case~\eqref{fFmod}, the two curves $W_1$ and $W_2$ are
distinguishable, since they wrap different cycles in the base, so we do
not have to be concerned with modding out by discrete symmetries.  

Since in our example all the curves are topologically spheres, there
are generically no vector fields in the low-energy theory. However, at
the points where there is a transition between the two-fivebrane
branch and the single fivebrane branch, additional low-energy fields
can appear. These correspond to membranes which stretch between the
two fivebranes becoming massless as the fivebranes
intersect.

\subsection{$[W]=\s_*l-\s_*E_1+F$}
\label{exbase2}

We can generalize the previous example by including a fiber component
in the class of $W$, so that 
\begin{equation}
   [W] = \s_*l - \s_*E_1 + F
\end{equation}
Note that from~\eqref{thrm} this class is effective. 

\begin{figure}
   \centerline{\psfig{figure=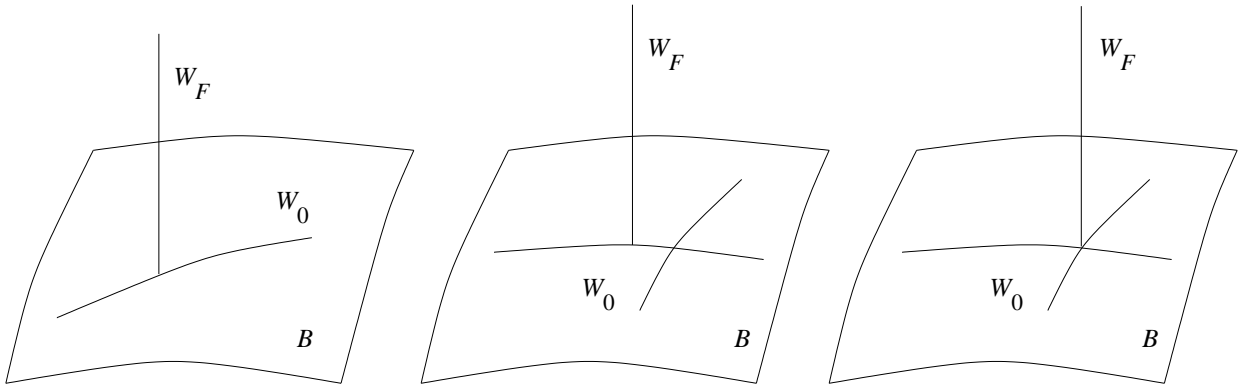,height=2in}}
   \caption{Possible enhancements in the moduli space of 
   $[W]=\s_*l-\s_*E_1+F$. The first figure is where the curve in the
   base $W_0$ does not split. The second gives two possible cases
   when $W_0$ splits.}
   \label{enhance}
\end{figure}

We immediately see that one simple possibility is that $W$ splits
into two curves 
\begin{equation}
   W = W_0 + W_F
\end{equation}
where 
\begin{equation}
   [W_0] = \s_*l - \s_*E_1, \qquad [W_F] = F
\end{equation}
The moduli space of the $W_0$ component will be exactly the same as
our previous example, while for the pure fiber component, as given in
equation~\eqref{Fmod}, the moduli space is locally $B\ts \PPa$. Since
the base in this example is $dP_8$ here we conclude that, when the
curve splits, this part of the moduli space is just the product of the
moduli spaces for $W_0$ and $W_F$, that is  
\begin{equation}
   \cM(\s_*l-\s_*E_1) \ts \left(dP_8 \ts \PPa\right)
\label{largecomp}
\end{equation}
where $\cM(\s_*l-\s_*E_1)$ was given above in~\eqref{lE1mod}. 
Physically we have two fivebranes, one wrapped on the fiber and one on
the base, which can each move independently. As discussed above, for
the curve wrapped on the base there are certain special points in
moduli space where it splits into a pair of fivebranes, so that, at
these special points, we have a total of three independent
fivebranes. Since the curves of the base are all topologically
spheres, their genus is zero. Hence, the only vector multiplets come
from the fivebrane wrapping the fiber which, being topologically a
torus with $g=1$, gives a $U(1)$ theory. Generically, the five brane
wrapping the fiber $W_F$ does not intersect the fivebranes in the base
$W_0$. However, there is a curve of points in the moduli space of
$W_F$ where both fivebranes are in the same position in $\orb$, with
the same value of $a$, and $W_0$ lies above $W_B$ in the Calabi--Yau
fibration. Generically, this gives a single intersection. However,
there is a special point, when the base curves splits, as
in~\eqref{simplesplit} and~\eqref{Wsplit}, and the fiber component
intersects exactly the point where the two base curves intersect. At
such a point in moduli space, we have three fivebranes intersecting at
a single point in the Calabi--Yau threefold. These different possible
intersections are shown in Figure~\ref{enhance}. Generically, we
expect there to be additional multiplets in the low-energy theory at
such points. 

We might expect that there is also a component of the moduli space
where the curve $W$ does not split at all, that is, where we have a
single fivebrane in the class $\s_*l-\s_*E_1+F$. To analyze this
second possibility we simply follow the analysis given above, where we
first consider the moduli space of the image $C$ of $W$ in the base
and then find the moduli space of curves which project down to the
same given curve $C$.

Since the projection of $F$ onto the base is zero, the image of $[W]$
in the base is $\O=l-E_1$, as above. Thus many of the results of
the previous discussion carry over to this situation. The moduli space
of $C$ is given by~\eqref{Cmodspace}. At a generic point in moduli
space $S_C=\p^{-1}(C)$ is a K3 surface, while at the seven special points
where $C$ splits, the surface above each of the two components of $C$
is a $dP_9$ surface. If we consider first a generic point in the
moduli space, $W$ is again a section of the K3 surface, but must now
be in the class $\s_*l-\s_*E_1+F$. How many such sections are there?
It turns out that for generic K3 there are none. We can see this as
follows. By adjunction, since the genus of $C$ was zero, we showed in
equation~\eqref{Omegasq1} that its class in $S_C$ satisfies $D_C\cdot
D_C=-2$. The identical calculation applies to any section, since
all sections
have 
genus zero. Thus, in particular, we have 
\begin{equation}
   [W]_C \cdot [W]_C = - 2
\label{contr1}
\end{equation}
where by $[W]_C$, we mean the class of $W$ in $S_C$. However, from
the map between classes~\eqref{K3inclmap}, it is clear that
$[W]_C=D_C+F_C$. Since $D_C$ is the class of the zero section, we
have $D_C\cdot F_C=1$. For the fiber class we always have
$F_C\cdot F_C=0$. Hence, we must also have
\begin{equation} 
   [W]_C \cdot [W]_C = 0
\label{contr2}
\end{equation}
This contradiction implies that there can be no sections of $S_C$ in
the class $\s_*l-\s_*E_1+F$. In other words, we have shown that,
generically, we cannot have just a single fivebrane in the class
$\s_*l-\s_*E_1+F$. Rather, the fivebrane always splits into a pure
fiber component and a pure base component, as described above. 

What about the special points in the moduli space where the curve $C$
splits into two? Do we still have to have a separate pure fiber
component? The answer is no, for the reason that, as discussed above,
the space above each component is a $dP_9$ surface and, unlike the K3
case, there are many more algebraic classes on $dP_9$ than just the
zero section and the fiber. Specifically, suppose there is no separate
pure fiber component in $W$ and consider the point where $C$ splits into
$C_1+C_2$ with $[C_1]=l-E_1-E_2$ and $[C_2]=E_2$. The actual curve $W$
must also split into $W_1$ and $W_2$. Given that each component must
be effective, we then have two possibilities, depending on which
component includes the fiber class 
\begin{equation}
\begin{gathered}
   {}[W_1] = \s_*l - \s_*E_1 - \s_*E_2 + F, \qquad  [W_2] = \s_*E_2 \\
   \text{or} \\
   {}[W_1] = \s_*l - \s_*E_1 - \s_*E_2, \qquad  [W_2] = \s_*E_2 + F
\end{gathered}
\label{dP9Fdecomp}
\end{equation}

\begin{figure}
   \centerline{\psfig{figure=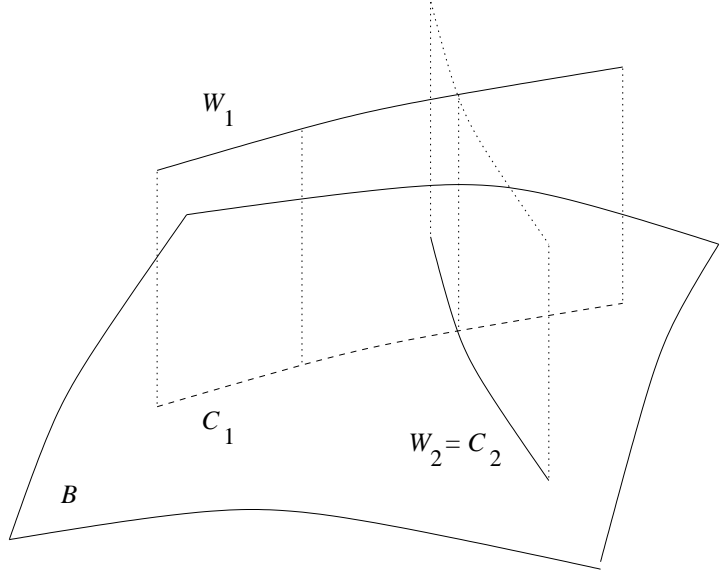,height=3in}}
   \caption{$W=W_1+W_2$ in the case where
   $[W_1]=\s_*l-\s_*E_1-\s_*E_2+F$ and $[W_2]=\s_*E_2$.}
   \label{discrete}
\end{figure}

Let us concentrate on the first case, although a completely analogous
analysis holds in the second example. Above, we calculated the number of
curves in the case where the class contains no $F$. We found,
for instance, that if $[W_2]=\s_*E_2$ then $W_2$ is required to be
precisely the section $C_2$ and there is no moduli space for moving
the curve in the fiber direction. The situation is richer, however,
for the class with an $F$ component. We will discuss this is more
detail in section~\ref{sec:dP9} below, but it turns out that there are
240 different sections of $dP_9$ in the class
$[W_1]=\s_*l-\s_*E_1-\s_*E_2+F$. It is a general result, just
repeating the calculation that led to~\eqref{dP9sec}, that any section
of the $dP_9$ has self-intersection $-1$. Consequently none of the 240
different sections in the class $\s_*l-\s_*E_1-\s_*E_2+F$ can be
deformed in the fiber direction and, hence, they simply provide a discrete
set of different $W_1$ which all map to the same $C_1$. Furthermore,
one can show that none of these sections intersect the base of the
Calabi--Yau manifold. Thus, since, in the case we are considering,
$W_2$ lies solely in the base, we find that $W_1$ and $W_2$ can never
overlap. Their relative positions within the Calabi--Yau threefold are
shown in Figure~\ref{discrete}. 

It is important to note that these curves are completely stuck within
the Calabi--Yau threefold. They cannot combine into a single curve and
move away from the exceptional point in the moduli space of $C$ (the
projection of $W$ into the $dP_8$ base) where $C$ splits into two
curves. Furthermore, we have argued that they cannot move in the
fiber. Thus, the only moduli for this component of the moduli space are
the positions of the two fivebranes in the orbifold interval and the
values of their axions, giving a moduli space of
\begin{equation}
   \PPa \ts \PPa
\label{smallcomp}
\end{equation}
Furthermore, since all the sections of $dP_9$ are topologically
spheres, there are no vector multiplets in this part of the moduli
space. As we noted above, the fivebranes cannot overlap. Hence, there
is no possibility of additional multiplets appearing. Finally, we note
that there were seven ways $C$ could split into $C_1+C_2$, and, for
each splitting, $W$ can decompose one of two
ways~\eqref{dP9Fdecomp}. Since for each decomposition there are 240
distinct sections, we see that there is a grand total of 3360 ways of
making the analogous decomposition to that we have just discussed. 

In conclusion, we see that the full moduli space for
$[W]=\s_*l-\s_*E_1+F$ has a relatively rich structure. It splits into
a large number of disconnected components. The largest component is
where $W$ splits into separate fiber and base components,
$W=W_0+W_F$. The moduli space is then given
by~\eqref{largecomp}, which includes the possibility of the base
component splitting. There are then 3360 disconnected components where
$W$ splits into two irreducible components, one of which includes the
fiber class $F$. We can summarize this structure in a table
\begin{equation}
\begin{array}{c|cc}
   [W]    & \text{genus} & \text{moduli space} \\
   \hline\hline
   (\s_*l-\s_*E_1) + (F) 
      & 0+1 
      & \left(\left[\PP^1 - 7\text{ pts.}\right]\ts\PPa\right)
          \ts \left(dP_8\ts\PPa\right) \\
   (\s_*l-\s_*E_1-\s_*E_2) + (\s_*E_2) + (F)
      & 0+0+1 & \PPa\ts\PPa \ts \left(dP_8\ts\PPa\right) \\
   \hline
   (\s_*l-\s_*E_1-\s_*E_2+F) + (\s_*E_2)
      & 0+0
      & \PPa\ts\PPa
\end{array}
\label{lE1Fmodspace}
\end{equation}
Here, the first column gives the homology classes in $X$ of the different
components of $W$. The first two rows describe the moduli
space~\eqref{largecomp} where $W$ splits into $W_0+W_F$, first for a
generic component in the base and then at one of seven points where
the base component splits. The final row describes one of the 3360
disconnected components of the moduli space~\eqref{smallcomp}. From
the genus count, we see that in the first two cases we expect a $U(1)$
gauge field on the fivebrane which wraps the fiber, while in the last
case there are no vector multiplets. As we have noted, the component
of the moduli given in the first two rows has the possibility that the
fivebranes intersect leading to additional low-energy fields, as
depicted in Figure~\ref{enhance}. The disconnected components, have no
such enhancement mechanism. Furthermore, their moduli space is
severely restricted since neither fivebrane can move within the
Calabi--Yau manifold.


\section{General procedure for analysis of the moduli space}
\label{gen}

From the examples above, we can distill a general procedure for the
analysis of the generic moduli space. We start with a general
fivebrane curve in the class
\begin{equation}
   [W] = \s_*\O + f F
\label{initialW} 
\end{equation}
Furthermore, $[W]$ is assumed to be effective, so that,
by~\eqref{thrm}, $\O$ is some effective class in $H_2(B,\ZZ)$ and
$f\geq 0$. We need to first find the moduli space of the projection
$C$ of $W$ onto the base. One then finds all the curves in the full
Calabi--Yau threefold in the correct homology class which project onto
$C$. In general, we will find that the above case, where the space
$S_C=\p^{-1}(C)$ above $C$ was a K3 surface, is the typical
example. There, we found that no irreducible curve which projects onto
$C$ could include a fiber component in its homology class. Hence, $[W]$
splits into a pure base and a pure fiber component. One exception, as
we saw above, is when $S_C$ is a $dP_9$ surface. In the following, we
will start by analysing the generic case and then give a separate
discussion for the case where $dP_9$ appears. 

\subsection{Decomposition of the moduli space}
\label{decomp}

Any curve $W$ in the Calabi--Yau threefold can be projected to the
base using the map $\pi$. In general, $W$ may have one or many
components. Typically, a component will project to a curve in the
base. However, there may also be components which are simply curves
wrapping the fiber at different points in the base. These curves will
all project to points in the base rather than curves. Thus, the first
step in analyzing the moduli space is to separate out all such
curves. We therefore write $W$ as the sum of two components, each of
which may be reducible,
\begin{equation}
   W = W_0 + W_F
\label{decompF}
\end{equation}
but with the assumption that none of the components of $W_0$ are pure
fiber components. For general $[W]$ given in~\eqref{initialW}, the
classes of these components are 
\begin{equation}
   [W_0] = \s_*\O + nF \qquad [W_F] = (f-n)F
\end{equation}
with $0\leq n \leq f$, since each class must be separately
effective. Note that, although $W_0$ has no components which are pure
fiber, its class may still involve $F$ since, in general, components of
$W_0$ can wrap around the fiber as they wrap around a curve in the
base. Except when $n=f$, so $W_F=0$, $W$ has at least two components in
this decomposition and so there are at least two five-branes. 

In general, the decomposition~\eqref{decompF} splits the moduli space
into $f+1$ different components depending on how we partition the $f$
fiber classes between $[W_0]$ and $[W_F]$. Within a particular
component, the moduli space is a product of the moduli space of $W_0$
and $W_F$. If $\cM_F((f-n)F)$ is the moduli space of $W_F$ and
$\cM_0(\s_*\O+nF)$ is the moduli space of $W_0$, we can then write the
full moduli space as
\begin{equation}
   \cM(\s_*\O + fF) = 
       \bigcup_{n=0}^f \cM_0(\s_*\O+nF) \ts \cM((f-n)F)
\label{moddecomp}
\end{equation}
The problem is then reduced to finding the form of the moduli spaces
for $W_0$ and $W_F$. The latter moduli space has already been analyzed
in section~\ref{sec:fF}. We found that, locally, 
\begin{equation}
   \cM((f-n)F) = \left(B\times\PPa\right)^{f-n}/\ZZ_{f-n}
\label{WFmod}
\end{equation}
Thus we are left with $\cM_0(\s_*\O+nF)$, which can be analyzed by
the projection techniques we used in the preceding examples. 

Projecting $W_0$ onto the base, we get a curve $C$ in the class $\O$ in
$B$. Let us call the moduli space of such curves in the base
$\cM_B(\O)$. This space is relatively easy to analyze since we know the
form of $B$ explicitly. In general, it can be quite complicated with
different components and branches as curves degenerate and split. 
To find the full moduli space $\cM_0(\s_*\O+nF)$, we fix a point in
$\cM_B(\O)$ giving a particular curve $C$ in $B$ which is the
projection of the original curve $W_0$ in the Calabi--Yau
threefold. In the following, we will assume that $C$ is not
singular. By this we mean that it does not, for instance, cross itself
or have a cusp in $B$. If it is singular, it is harder to analyze the
space of curves $W_0$ which project onto $C$. In general, $C$ splits
into $k$ components, so that
\begin{equation}
   C = C_1 + \dots + C_k
\label{genCdecomp}
\end{equation}
We then also have
\begin{equation}
   \O = \O_1 + \dots + \O_k
\end{equation}
where $\O_i=[C_i]$ and, so, must be an effective class on the base for
each $i$. Clearly if the curve in the base has more than one component
then so does the original curve $W_0$, so that 
\begin{equation}
   W_0 = W_1 + \dots + W_k
\label{Widecomp}
\end{equation}
with $\p(W_i)=C_i$. In general, the class of $W_0$ will be partitioned
into a sum of classes of the form
\begin{equation}
   [W_i] = \s_*\O_i + n_i F
\label{Wiclasses}
\end{equation}
where, for each curve to be effective, $n_i\geq 0$ and
$n_1+\dots+n_k=n$, leading to a number of different possible
partitions. 

One now considers a particular component $C_i$. To find the moduli
space over $C_i$, one needs to find all the curves $W_i$ in $X$ in
the cohomology class $\s_*\O_i+n_iF$ which project on $C_i$. Recall,
in addition, that we have assumed in our original
partition~\eqref{decompF} that $[W_i]$ contains no pure fiber
components. Repeating this procedure for each component and for each
partition of the $n$ fibers into $\{n_i\}$, gives the moduli space over a
given point $C$ in $\cM_B(\O)$ and, hence, the full moduli space. 

Consequently, we have reduced the problem of finding the full moduli
space $\cM_0(\s_*\O+nF)$ to the following question. To simplify
notation, let $R$ be the given irreducible curve $C_i$ in the base,
and let $V$ be the corresponding curve $W_i$ in the full Calabi--Yau
threefold. Let us further write the class $\O_i$ of $C_i$ as $\L$ and
write $m$ for $n_i$. Our general problem is then to find, for the
given irreducible curve $R$ in the effective homology class $\L$, what
are all the curves $V$ in the Calabi--Yau threefold in the class
$\s_*\L+mF$, where $m\geq 0$ which project onto $R$. Necessarily, all
the curves $V$ which project into $R$ lie in the surface
$S_R=\p^{-1}(R)$. By construction, $S_R$ is elliptically fibered over
the base curve $R$. Furthermore, typically, the map from $V$ to $R$
wraps $R$ only once. It is possible that $R$ is some number $q$ of
completely overlapping curves in $B$, so that $[R]=q\G$ for some
effective class $\G$ in $B$. Then the map from $V$ to $R$ wraps the
base curve $q$ times. This will occur in one of the examples we give
later in the paper, but here, since we are discussing generic
properties, let us ignore this possibility. Then, assuming $R$ is not
singular, the map is invertible and we see that $V$ must be a section
of the fibered surface $S_R$. 

Furthermore, we note that $S_R$ can be characterized by the genus $g$
of the base curve $R$ and the number of singular fibers. The former
is, by adjunction, given by
\begin{equation}
   2g - 2 = \left(K_B + \L\right) \cdot \L
\label{genus}
\end{equation}
The latter is also a function only of the class $\L$ of $R$ and can be
found by intersecting the discriminant class $[\D]$ with $\L$. Since
$[\D]=-12K_B$, the number of singular fibers must be of the form $12p$
with
\begin{equation}
   p = - K_B \cdot \L
\label{pdef}
\end{equation}
where $p$ is an integer. If $p$ is negative, the curve $R$ lies
completely within the discriminant curve of the elliptically
fibered Calabi--Yau manifold. This means that every fiber above $R$ is
singular. The form of $S_R$ then depends on the structure of the
particular fibration of the Calabi--Yau threefold. Since we want to
consider generic properties of the moduli space, we will ignore this
possibility and restrict ourselves to the case where $p$ is non-negative. We
note that this is not very restrictive. For del Pezzo and Enriques
surfaces $-K_B$ is nef, meaning that its intersection with any
effective class in the base is non-negative. Hence, since $\L$ must be
effective, we necessarily have $p\geq 0$. The only exceptions, are
Hirzebruch surfaces $F_r$ with $r\geq 3$ and where $\L$ includes the
negative section $\cS$.  

Finally, then, finding the full moduli space $\cM_0(\s_*\O+nF)$ has
been reduced to the following problem 
\begin{itemize}
\item
For a given irreducible curve $R$ in the base $B$ with homology class
$\L$, find the moduli space of sections $V$ of the surface
$S_R=\p^{-1}(R)$ in the homology class $\s_*\L+mF$ in the full
Calabi--Yau threefold $X$, where $m\geq 0$. 

$S_R$ is characterized by $g=\genus{C}$, as given in~\eqref{genus},
and $p$, where $12p$ is the number of singular elliptic fibers, as
given in~\eqref{pdef}. Consequently, we write this moduli space as
$\cM(g,p;m)$. 
\end{itemize}
We will assume that $p\geq 0$. This is necessarily true, except when
$B$ is an $F_r$ surface with $r\geq 3$ and $\L$ contains the class of
the section at infinity $\cS$.

\subsection{The generic form of $\cM(g,p;m)$}
\label{generic}

To understand the sections of $S_R$, we start by finding the algebraic
classes on $S_R$. As for the K3 surface, a generic surface $S$ has no
algebraic classes since $h^{2,0}\neq 0$. For $S_R$, we know that two
classes are necessarily present, the class of the zero section $D_R$
and the fiber class $F_R$. However, generically, there need not be any
other classes. Additional classes may appear for special choices of
complex structure, but here we will consider only the generic
case. The obvious exception is the case where $g=0$ and $p=1$. From
equations~\eqref{genus} and~\eqref{pdef}, this implies that $R$ is an
exceptional curve in $B$. 
The surface $S_R$ is then an elliptic fibration over $\PP^1$ with 12
singular fibers, which is a $dP_9$ surface. In this case $h^{2,0}=0$
and every class in $H_2(S_R,\ZZ)$ is algebraic. We will return to this
case in the next section. The inclusion map $i_R:S_R\to X$, gives a
natural map between classes in $S_R$ and in $X$
\begin{equation}
   i_{R*} : H_2(S_R,\ZZ) \to H_2(X,\ZZ)
\end{equation}
In general, with only two classes the map is simple. By construction,
we have
\begin{equation}
\begin{aligned}
   i_{R*}D_R &= \s_*\L \\
   i_{R*}F_R &= F
\end{aligned}
\label{geninclmap}
\end{equation}

Just as in the K3 example, we will find that the existence of only two
classes strongly constrains the moduli space $\cM(g,p;m)$. We can find
the 
analog 
of the contradiction of equations~\eqref{contr1}
and~\eqref{contr2} as follows. Let $K_{S_R}$ be the cohomology class
of the canonical bundle of the surface. Since both $V$ and $R$ are
sections, they have the same genus $g$. By the Riemann--Hurwitz
formula and adjunction, we have
\begin{equation}
   2g - 2 = \left(K_{S_R} + D_R\right) \cdot D_R 
          = K_{S_R} \cdot D_R + D_R \cdot D_R 
\label{Radj}
\end{equation}
where $D_R$ is the class of the zero section, and
\begin{equation}
   2g - 2 = \left(K_{S_R} + [V]_R\right) \cdot [V]_R 
          = K_{S_R} \cdot [V]_R + [V]_R \cdot [V]_R 
\label{Vadj}
\end{equation}
where $[V]_R$ the class of $V$ in $S_R$. We also have, for the fiber,
by a similar calculation, since it has genus one and, since two
generic fibers do not intersect, $F_R\cdot F_R=0$, that 
\begin{equation}
   0 = \left(K_{S_R} + F_R\right) \cdot F_R = K_{S_R} \cdot F_R 
\end{equation}
Finally, since we require in $X$ that $[V]=\s_*\L+mF$ and since we
assume that no additional classes exist on $S_R$, %
\begin{equation}
   [V]_R = D_R + mF_R
\label{Vclass}
\end{equation}
Substituting this expression into~\eqref{Vadj} and using~\eqref{Radj},
we find 
\begin{equation}
   [V]_R \cdot [V]_R = D_R \cdot D_R
\label{Vsqr1}
\end{equation}
On the other hand, given that $F_R\cdot D_R=1$ because the fiber
intersects a section at one point, we can compute the
self-intersection of $[V]_R=D_R+mF_R$, explicitly, yielding
\begin{equation}
   [V]_R \cdot [V]_R = D_R \cdot D_R + 2m
\label{Vsqr2}
\end{equation}
Comparing~\eqref{Vsqr1} with~\eqref{Vsqr2}, we are left with the
important conclusion that we must have $m=0$. This implies that,
generically, no component of $W_0$ can contain any fibers in its
homology class. We conclude that  
\begin{equation}
   \cM(g,p;m) = \emptyset \qquad \text{unless} \quad m=0
\end{equation}

In fact, we can go further. Since we require $m=0$, we see
from~\eqref{Vclass} that $V$ is in the same class $D_R$ as the zero
section $R$. Using the 
Riemann--Roch formula and Kodaira's description of elliptically
fibered surfaces, one can show that the canonical class in the
cohomology of $S_R$ is
given by 
\begin{equation}
   K_{S_R} = \left(2g - 2 + p\right) F_R
\label{KSR}
\end{equation}
Then, substituting this expression into~\eqref{Radj}, we see that 
\begin{equation}
   D_R \cdot D_R = - p 
\end{equation}
Thus we see that for $p>0$, $R$ is an exceptional divisor and cannot
be deformed in $S_R$. Consequently, all the fivebrane can do is to
move in the orbifold direction and change its value of $a$, so we have a
moduli space of $\PPa$. If $p=0$, the fibration is locally
trivial. It may or may not be globally trivial. However, it is always
globally trivial when pulled back to some finite cover of the
base. Every section is in the class $D_R$ and the moduli space simply
corresponds to moving the fivebrane in the fiber direction and in
$\orb$ and $a$. If the original fibration was globally trivial, this
yields a moduli space of $E\ts\PPa$, where $E$ is an elliptic curve
describing motion of $V$ in the fiber direction. If it was only
locally trivial, then these deformations in the fiber directions still
make sense over the cover, but only a finite subset of them happens to
descend, so the actual moduli space consists in this case of some
finite number of copies of $\PPa$. 

In summary, we see that, generically, if we exclude the case $g=0$,
$p=1$ where $S_R$ is a $dP_9$ surface, 
\begin{equation}
   \cM(g,p;m) = \emptyset \qquad \text{for} \quad m > 0 
\label{gencMR1}
\end{equation}
while if $m=0$ we have, 
\begin{equation}
   \cM(g,p;0) = 
     \begin{cases} 
        N \PPa & \text{if } p = 0 
           \text{ and the fibration is not globally trivial} \\
        E\ts\PPa & \text{if } p = 0 
           \text{ and the fibration is globally trivial}\\
        \PPa & \text{if } p > 0 
     \end{cases}
\label{gencMR2}
\end{equation}
where $N$ is some integer depending on the global structure of $S_C$.
In each case, the gauge group on the fivebrane is given by $U(1)^g$
where $g$ is the genus of the curve $R$ in the base $B$. Since both
$R$ and $V$ are sections of $S_R$, this is equal to the genus of the
curve $V$ in the space $X$.

\subsection{The $dP_9$ exception and ${\cal{M}}(0,1;m)$}
\label{sec:dP9}

As we have mentioned, the obvious exception to the above analysis is
when $S_R$ is a $dP_9$ surface. This occurs when the base curve $R$ is
topologically $\PP^1$ and there are 12 singular elliptic fibers in
$S_R$, that is if $[R]=\L$ in the base $B$, 
\begin{equation}
   2g - 2 = \left(K_B+\L\right)\cdot\L = -2, \qquad 
   p = - K_B \cdot \L = 1
\label{dP9cond}
\end{equation}
As we mentioned above, this implies that $R$ is an exceptional curve
in the base. 
In this case, $D_R$ and $F_R$ are not the only generic algebraic
classes on $S_R$. Rather, since $h^{2,0}=0$, every integer class in
$dP_9$ is algebraic. The surface $dP_9$ can be described as the plane
$\PP^2$ blown up at nine points which are at the intersection of two
cubic curves. Consequently, there are ten independent algebraic
classes on $dP_9$, the image $l'$ of the class of a line in $\PP^2$
and the nine exceptional divisors, $E'_i$ for $i=1,\dots,9$,
corresponding to the blown up points. Here we use primes to
distinguish these classes from classes in the base $B$ of the
Calabi--Yau threefold (specifically, the $E_i$ classes in $B$ when the
base is a del Pezzo surface).

The point here is that, in the full Calabi--Yau manifold $X$, there
are only two independent classes associated with $S_R$, namely, the
class of the base curve $\s_*\L$ and the fiber class
$F$. Consequently, if $i_R:S_R\to X$ is the inclusion map from the
$dP_9$ surface into the Calabi--Yau threefold, the corresponding map
between classes 
\begin{equation}
   i_{R*} : H_2(S_R,\ZZ) \to H_2(X,\ZZ)
\label{iRmap}
\end{equation}
must have a non-empty kernel, since it maps the ten independent
classes on $S_R$ mentioned above into only two in $X$. Recall that our
goal is to find the set of sections $V$ of $S_R$ which are in the
class $\s_*\L+mF$ in $X$. Previously, we found that generically there
were no such sections unless $m=0$. This followed from
equations~\eqref{Radj} to~\eqref{Vsqr2}. Now, the appearance of a
kernel in the map~\eqref{iRmap}, means that $[V]_R$ is no longer
necessarily of the form $D_R+mF_R$, as in
equation~\eqref{Vclass}. Consequently we can no longer conclude that
we must have $m=0$. In fact, as we will see, there are several
different sections $V$ in the same class $\s_*\L+mF$ in $X$ with
$m>0$.

\begin{figure}
   \centerline{\psfig{figure=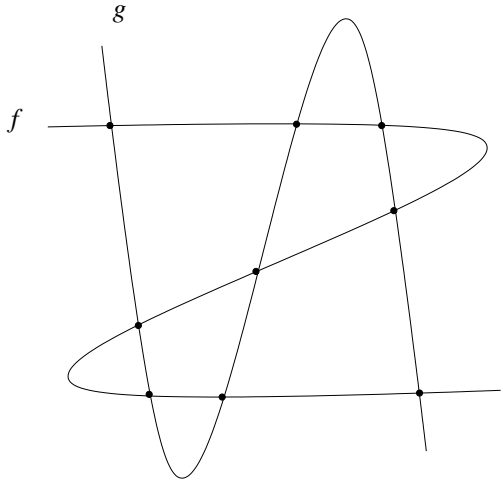,height=2.5in}}
   \caption{The space $dP_9$}
   \label{dP9}
\end{figure}

Let us start by recalling why a $dP_9$ surface is also an elliptic
fibration. Viewed as $\PP^2$ blown up at nine points, we have the
constraint that the nine points lie at the intersection of two
cubics. This is represented in Figure~\ref{dP9}. The cubics can be
written as two third-order homogeneous polynomials in homogeneous
coordinates $[x,y,z]$ on $\PP^2$. Let us call these polynomials $f$
and $g$. Clearly any linear combination $af+bg$ defines a new cubic
polynomial. By construction, this polynomial also passes through the
same nine points. Since the overall scale does not change the cubic,
the set of cubics is given by specifying $a$ and $b$ up to overall
scaling. Thus, we have a $\PP^1$ of cubics passing through the nine
points. Since each cubic defines an elliptic curve, we can think of
this as an elliptic fibration over $\PP^1$. Furthermore, the cubics
cannot intersect anywhere else in $\PP^2$. The space of cubics spans
the whole of the plane and, further, it blows up each intersection point into a
$\PP^1$ of distinct points, one point in $\PP^1$ for each cubic
passing through the intersection. Thus the space of cubics is giving
an alternative description of the $dP_9$ surface. In addition, we note
that each of the exceptional divisors $E'_i$, the blow-ups of the
intersection points, is a $\PP^1$ surface which intersects each fiber
at a single point, and so corresponds to section of the
fibration. Furthermore, the anti-canonical class of $dP_9$ is given by
\begin{equation}
   - K_{S_R} = 3l' - E'_1 - \dots - E'_9 = F_R
\label{dP9FR}
\end{equation}
and is precisely the class of the cubics passing through the nine
intersection points. It follows that 
\begin{equation}
   K_{S_R} = - F_R
\end{equation}
which, since $g=0$ and $p=1$, agrees with the general
expression~\eqref{KSR}. 

It is natural to ask if there are other sections of $dP_9$ aside from
the exceptional curves $E'_i$. We first note that any section of
$dP_9$ is exceptional, with $[V]_R\cdot[V]_R=-1$. This follows
from~\eqref{Vadj} and~\eqref{dP9FR}, recalling that any section will 
have genus zero and intersects the fiber (the anti-canonical class)
only once. Next, we recall that there is a notion of addition of points
on elliptic curves. Consequently, we can add sections point-wise to get
a new section. Thus, we see that the set of sections forms an infinite
Abelian group containing all the exceptional curves on the $dP_9$
surface. From the point of view of curves in $\PP^2$, these additional
exceptional classes correspond to curves of higher degree passing
through the nine intersection points with some multiplicity (see for
instance~\cite{dlow2}). In general, we can write an exceptional class
(except the classes $E'_i$) as 
\begin{equation}
   Q = q l' - \sum q_i E'_i 
\label{gensec}
\end{equation}
such that 
\begin{equation}
   q, q_i \geq 0, \quad
   q^2 - \sum q_i^2 = -1, \quad
   3q - \sum q_i = 1
\label{genseccond}
\end{equation}
where the first condition is required in order to describe an
effective curve in $\PP^2$, the second condition gives $Q\cdot Q=-1$ and
the third condition gives $Q\cdot F_R=1$. The appearance of an infinite
number of exceptional classes means that these equations have an
infinite number of solutions for $i=1,\dots,9$. 

Having identified all the relevant classes in $dP_9$, we can now turn
to a description of the map~\eqref{iRmap}. First, we note that the
group structure of the set of sections means that we can choose any
section as part of the basis of classes. In our case, we have singled
out one section as the class of the base curve $R$. Thus, without loss
of generality, we can identify this class with one of the $E'_i$, for
example, $E'_9$. Thus we set
\begin{equation}
   D_R = E'_9
\end{equation}
By construction, we know that $D_R$ maps to $\s_*\L$ and $F_R$ maps to
$F$ under $i_{R*}$. Thus, in terms of $l'$ and $E'_i$, we have,
using~\eqref{dP9FR} and the generic result~\eqref{geninclmap}, 
\begin{equation}
\begin{gathered}
   i_{R*} E'_9 = \s_*\L \\
   i_{R*} \left(3l'-E'_1-\dots-E'_9\right) = F
\end{gathered}
\label{genincl}
\end{equation}
We now need to understand how the remaining independent classes
$E'_1,\dots,E'_8$ map under $i_{R*}$. We first note that, since all
these classes are sections, they must project onto the class $\L$ in
the base. The only ambiguity is how many multiples of the fiber class
$F$ each contains. Thus, we know 
\begin{equation}
   i_{R*}E'_i = \s_*\L + c_i F
\end{equation}
for some $c_i$. The easiest way to calculate $c_i$ is to recall that a
class in $H_2(X,\ZZ)$ is uniquely determined by its intersection
numbers with a basis of classes in $H_4(X,\ZZ)$. A suitable basis was
given in section~\ref{algclasses}. In particular, we can consider the
intersection of $E'_i$ with the class $D$ of base $B$ of the full
Calabi--Yau threefold. We note that $E'_i\cdot E'_9=0$ for
$i=1,\dots,8$, so the extra sections $E'_i$ do not intersect the base
$R$ of the $dP_9$. However, since $R$ describes the intersection of
$B$ with the $dP_9$ surface $S_R$, we see that the extra classes
cannot intersect $B$. Thus, we must have
\begin{equation}
   i_{R*} E'_i \cdot D = 0 \qquad \text{for} \quad i = 1,\dots,8
\end{equation}
From the table of intersections~\eqref{inters}, using~\eqref{dP9cond},
we see that we must have $c_i=1$ for $i=1,\dots,8$. Together with the
result~\eqref{genincl} we find, in conclusion, that the
map~\eqref{iRmap} is given by 
\begin{equation}
\begin{aligned}
   i_{R*} l' &= 3 \s_*\L + 3 F \\
   i_{R*} E'_i &= \s_*\L + F \qquad \text{for} \quad i = 1,\dots,8 \\
   i_{R*} E'_9 &= \s_*\L
\end{aligned}
\label{fulliR}
\end{equation}
demonstrating explicitly that the map has a kernel. 

Having identified the map, we would now like to return to our original
question, which was how many sections are there in the class
$\s_*\L+mF$ and what is their moduli space. The second part is easy
to answer. We have noted that all sections are exceptional with
self-intersection $-1$. This implies that they cannot be moved within
the $dP_9$ surface. However, they can move in the orbifold and have different
values of the axion. Consequently, each section has a moduli space of
$\PPa$. If there are a total of $N(m)$ sections for a given $m$,
then the total moduli space $\cM(0,1;m)$ has the form 
\begin{equation}
   \cM(0,1;m)=N(m)\PPa
\label{cMm}
\end{equation}
The value of $N(m)$ is a problem in
discrete mathematics. Recall that the class of a general section had
the form given in~\eqref{gensec}. Under the map $i_{R*}$ this maps
into 
\begin{equation}
   i_{R*}Q = \s_*\L + \left(q_9+1\right) F
\end{equation}
where we have used the condition $3q-\sum q_i=1$. Thus, we can
summarize the problem as finding the number of solutions to 
\begin{equation}
\begin{aligned}
   q^2 - \sum q_i^2 + 1 &= 0 \\
   3q - \sum q_i &= 1
\end{aligned}
\label{sectcond}
\end{equation}
with 
\begin{equation}
   q \geq 0, \quad q_i \geq 0, \quad q_9 = m - 1
\end{equation}

We will not solve this problem in general, but just note the solution
of $m=0$ and $m=1$. The former case is not actually included in the
above form, since it implies $q_9=-1$. However, this case is easy to
analyze since, with $m=0$, we see, from~\eqref{inters}
and~\eqref{dP9cond}, that the intersection of $[V]=\s_*\L$ with the
base class $D$ of the Calabi--Yau manifold is $-1$. Consequently, a
component of $V$ must lie in the base $B$. Since, by assumption, $V$ is
irreducible, this means that the whole of $V$ lies in $B$. Thus, $V$ can
only be the base section $R$ in the $dP_9$ surface. Thus we conclude that,
for a $dP_9$ surface $S_R$, $N(0)=1$. That is,
\begin{equation}
   \cM(0,1;0) = \PPa
\end{equation}
corresponding to moving the single curve $R$ in $\orb$. This
result was used in analyzing the example in section~\ref{exbase1}. 

For $m=1$, there are 240 solutions to the equations~\eqref{sectcond},
that is $N(1)=240$. 
The easiest way to see this is to
note that $m=1$ implies $q_9=0$. Thus, we can ignore
$E'_9$. Effectively, one is then trying to count the number of
exceptional curves on a $dP_8$ surface. This is known to be a finite
number, $240$~\cite{dPref}. Explicitly, they are of the following
forms
\begin{equation}
\begin{aligned}
   {}& E'_i \\
   {}& l' - E'_i - E'_j \\
   {}& 2l' - E'_{i_1} - \dots - E'_{i_5} \\
   {}& 3l' - 2E'_{i} - E'_{i_1} - \dots - E'_{i_6} \\
   {}& 4l' - 2E'_{i_1} - \dots - 2E'_{i_3} - E'_{i_4} - \dots - E'_{i_8} \\
   {}& 5l' - 2E'_{i_1} - \dots - 2E'_{i_6} - E'_{i_7} - E'_{i_8} \\
   {}& 6l' - 2E'_{i_1} - \dots - 2E'_{i_7} - 3E'_{i_8}
\end{aligned}
\end{equation}
where all the indices run from $1$ to $8$. One can see, 
using~\eqref{fulliR}, that all these classes map under $i_{R*}$ to
$\s_*\L+F$. Again, since none of these sections can move in the $dP_9$
surface, the moduli space is simply 240 copies of $\PPa$, 
\begin{equation}
   \cM(0,1;1) = 240\, \PPa
\end{equation}
This result was used in analyzing the example in
section~\ref{exbase2}. We also note that none of these sections
intersects the base section $R$. 


\section{A three-family $dP_8$ example}

In the previous section we have given a general procedure for
analyzing the moduli space of fivebranes wrapped on a holomorphic
curve in a particular effective class in the Calabi--Yau manifold. The
problem is of particular importance because it has been
shown~\cite{nse,dlow1,dlow2} that including fivebranes in
supersymmetric M~theory compactifications greatly enlarges the number
of vacua with reasonable grand unified gauge groups and three families
of matter.

In the remaining sections, we apply this procedure to some
specific examples which arise in the construction of phenomenological
models. We will find that the moduli spaces are very
rich. Nonetheless, we will find that there are isolated
parts of moduli space where the number of moduli is greatly
reduced. We will not describe the full moduli spaces here, but rather
consider various characteristic components. 

Let us start with the example given in~\cite{dlow1} and expanded upon
in~\cite{dlow2}. There, the base $B$ was a $dP_8$ surface and the class
of the fivebranes was given by 
\begin{equation}
   [W]= 2\s_*E_1 + \s_*E_2 + \s_*E_3 + 17F
\label{eq:y1}
\end{equation}
where $l$ and $E_i$ for $i=1,\dots,8$ are the line class and the
exceptional blow-up classes in the $dP_8$. Since $2E_1+E_2+E_3$ is
effective in $dP_8$, this describes an effective class in the
Calabi--Yau manifold. This choice of fivebrane class, together with a
non-trivial $E_8$ bundle $V_1$, led to a low-energy $SU(5)$ theory
with three families.

\subsection{General decomposition}

Let us follow exactly the procedure laid down in the previous
section. First, we separate from $W$ all the pure fiber components,
writing it as the sum of $W_0$ and $W_F$ as in~\eqref{decompF}. We
write
\begin{equation}
   [W_0] = 2E_1 + E_2 + E_3 + nF, \qquad [W_F] = (17-n) F
\end{equation}
with $0\leq n\leq 17$. Unless $n=17$, we have at least two
separate fivebranes. This splits the moduli space into several
different components depending on the partition of 17 into $n$ and
$17-n$, as given in~\eqref{moddecomp}. The moduli space for the pure
fiber component $W_F$ is just the familiar form, read off
from~\eqref{WFmod} 
\begin{equation}
   \cM((17-n)F) = \left(dP_8 \ts \PPa\right)^{17-n}/\ZZ_{17-n}
\end{equation}

More interesting is the analysis of the $W_0$ moduli space. As
described above, the first step in the analysis is to project $W_0$
onto the base. This gives the curve $C$ which is in the homology
class
\begin{equation}
   [C] = 2E_1 + E_2 + E_3
\end{equation}
We then need to find the moduli space $\cM_B(2E_1+E_2+E_3)$ of $C$ in
the base. We recall that the del Pezzo surface $dP_8$ can be viewed as
$\PP^2$ blown up at eight points. The exceptional classes $E_i$ each
have a unique representative, namely the exceptional curve $\PP^1$ at the
$i$-th blown-up point. Furthermore, we have the intersection numbers
$E_i\cdot E_j=-\d_{ij}$. Thus $[C]$ has a negative intersection number
with each of $E_1$, $E_2$ and $E_3$. This implies that it must have a
component contained completely within each of the exceptional curves
described by $E_1$, $E_2$ and $E_3$. Since these are all distinct, $C$
must be reducible into three components 
\begin{equation}
   C = C_1 + C_2 + C_3
\end{equation}
where
\begin{equation}
   [C_1] = 2E_1 \qquad 
   [C_2] = E_2 \qquad
   [C_3] = E_3
\end{equation}
None of these components can be moved in the base, since they all have
negative self-intersection number. Consequently, we have 
\begin{equation}
   \cM_B(2E_1+E_2+E_3) = \text{single pt.}
\end{equation}
corresponding to three fivebranes, each wrapping a different
exceptional curve.

Since the projection $C$ splits, so must the curve $W_0$ itself. We
must have
\begin{equation}
   W_0 = W_1 + W_2 + W_3
\end{equation}
We can partition the fiber class in $[W_0]$ in different ways. In
general we write 
\begin{equation}
   [W_1] = 2\s_*E_1 + n_1 F \qquad
   [W_2] = \s_*E_2 + n_2 F \qquad
   [W_3] = \s_*E_3 + n_3 F
\label{split1}
\end{equation}
with $n_1+n_2+n_3=n$ and $n_i\geq0$ since each curve must be
separately effective. The problem of finding the full moduli space has
now been reduced to finding the moduli space of $W_1$, $W_2$ and $W_3$
separately. 

We see that, unless $n=17$, we have at least four separate
five-branes, one wrapping the pure fiber curve $W_F$ and one wrapping
each of the three curves $W_1$, $W_2$ and $W_3$. As discussed in
section~\ref{sec:fF}, the pure fiber component can move in the base
$B$, as well as in the orbifold. Furthermore it has transitions where
it separates into more than one fivebrane. The components $W_i$,
meanwhile, are stuck above fixed exceptional curves in the base. They
are free to move in the orbifold and may be free to move in the fiber
direction (this will be discussed in the following sections). The
$W_i$ components cannot intersect since the exceptional curves in the base over
which they are stuck cannot intersect. However, the pure fiber
components can intersect the $W_i$, leading to the possibility of
additional low-energy fields appearing.

\subsection{The $W_2$ and $W_3$ components}
\label{W2W3}

We start by analyzing the $W_2$ and $W_3$ components. As usual, we are
interested in the number of sections of the surface $S_{C_i}$ which
are in the class $\s_*E_i+n_iF$ in the full Calabi--Yau manifold for
$i=2,3$. In each case, the base curve $C_i$ wraps an exceptional
curve $\PP^1$ of one of the blown-up points in $dP_8$. Such a case is
familiar from the examples given in sections~\ref{exbase1}
and~\ref{exbase2}. The corresponding surface $S_{C_i}$, as expected,
since $C_i$ wraps an exceptional curve, is a $dP_9$
manifold. Explicitly, for both $C_2$ and $C_3$ we have 
\begin{equation}
   g_i = \text{genus}(C_i) = 0
\end{equation}
and the number of singular fibers in the fibration is given by $12p_i$, where
\begin{equation}
\begin{split}
   p_i &= - K_{dP_8} \cdot E_i \\
     &= \left(3l-E_1-\dots-E_8\right)\cdot E_i = 1
\end{split}
\end{equation}
Thus, we see that 
\begin{equation}
   S_{C_i} \text{ is a $dP_9$ surface for } i = 2, 3
\end{equation}
The moduli space for each $W_{i}$, for $i=2,3$, is then the moduli space of
sections of the $dP_{9}$ surface in the homology class $\s_*E_i + n_i F$ in
the full Calabi--Yau space $X$. Since in this case $g_{i}=0$ and $p_{i}=1$, it
follows that we are interested in the moduli spaces $\cM(0,1;n_i)$
discussed in section~\ref{sec:dP9}. 

Let us concentrate on $W_2$, since the moduli space of $W_3$ is
completely analogous. We recall from section~\ref{sec:dP9} that, for a
given $n_2$, there are a finite number of sections of the $dP_9$ in the
class $\s_*E_2 + n_2 F$ in $X$. Furthermore, all these sections are exceptional
with self-intersection $-1$ and so there is no moduli space for
moving these curves within $dP_{9}$. Since the curve $C_2$ is also
fixed in the base, we see that there are no moduli for moving $W_2$
in the Calabi--Yau threefold. All we are left with are the moduli for
moving in $\orb$ and the axion modulus. 

We showed in section~\ref{sec:dP9} that there was precisely one section
in the class $\s_*E_2$ and 240 in $\s_*E_2+F$. Rather than do a
general analysis, let us consider some examples for which
$n_2=2$. That is
\begin{equation}
   [W_2]=\s_*E_2 + 2F
\label{eq:hello1}
\end{equation}
Consequently, we will be interested in the moduli space $\cM(0,1;2)$. We know
from the previous discussion that
\begin{equation}
   \cM(0,1;2)=N(2)\PPa
    \label{eq:hi}
\end{equation}
Here, we will not evaluate $N(2)$ but content ourselves with several specific
examples.
From the general map~\eqref{fulliR}, with $\L=E_2$, we see
that we can, for instance, write the class of $W_2$ as
\begin{equation}
   [W_2] = i_{C_2*}\left(l'-E'_1-E'_9\right)
   \label{eq:hello2}
\end{equation}
or
\begin{equation}
   [W_2]= i_{C_2*}\left(2l'-E'_1-E'_2-E'_3-E'_4-E'_9\right)
   \label{eq:hello3}
\end{equation} 
or many other decompositions which we will not discuss here.
We might be tempted to include the case where
\begin{equation}
   [W_2] = i_{C_2*}\left(3l' - E'_2 - \dots - E'_9\right)
\label{badex}
\end{equation}
However, this is not a section. This can be seen directly from the
fact that it fails to satisfy the
equations~\eqref{sectcond}. Alternatively, we note that the nine blown
up points in $dP_9$ are not in general position. If a cubic passes
through eight of them, then it also passes through the
ninth. Consequently, the class $3l'-E'_2-\dots-E'_9$, which is the
class of a cubic through eight of the nine points, always splits into
two classes: the fiber $F_{C_2}=3l'-E'_1-\dots-E'_9$ and the base
$E'_1$. Consequently, the example~\eqref{badex} is always reducible,
splitting into a pure fiber component and the section in the class
$E'_1$. 

Since all the sections are genus zero like the base, we have a simple
table for the cases given in~\eqref{eq:hello2} and~\eqref{eq:hello3}
\begin{equation}
\begin{array}{c|cc}
   [W_2] \text{ in $dP_9$} & \text{genus} & \text{moduli space} \\
   \hline
   l'-E'_1-E'_9 & 
      0 & \PPa \\
   2l'-E'_1-E'_2-E'_3-E'_4-E'_9 &
      0 & \PPa 
\end{array}
\end{equation}
All the other possible sections in the class $\s_*E_2+2F$ will have
the same genus and moduli space.

\subsection{The $W_1$ Component}

The remaining component $W_1$ is considerably more interesting. We
start by noting that the projection $C_1=2R$, where $R$ is the
exceptional curve corresponding to $E_1$. This implies that the
projection map from $W_1$ to $R$ is a double cover. For the same
reason as in the previous section, we have that the fibered space
$S_R$ above $R$ satisfies
\begin{equation}
   S_R \text{ is a $dP_9$ surface}
\end{equation}
However, the fact that $W_1$ is a double cover implies that, unlike
all the cases we have considered thus far, $W_1$ is not a
section. Hence, the $W_{1}$ moduli space is not given by $\cM(0,1;n_{1})$.
However, we can still analyze its moduli space using
the general map between classes~\eqref{fulliR}. 

Suppose, for simplicity, we choose $n_1=2$, that is
\begin{equation}
 [W_1] = 2E_1 + 2 F 
\label{eq:today1}
\end{equation}
since other cases can be
analyzed in an analogous way. As above, we will not consider all
possibilities for the class of $W_1$ in the $dP_9$. Instead, we will restrict
our discussion to a few interesting examples.
For instance,
using~\eqref{fulliR} with $\L=E_1$, some possibilities are
\begin{subequations}
\label{W1ex}
\begin{align}
   {} [W_{1}]&= i_{R*}\left(2E'_1\right)               \label{W1ex1}\\
   {} [W_{1}]&= i_{R*}\left(E'_1+E'_2\right)           \label{W1ex2}\\
   {} [W_{1}]&= i_{R*}\left(l'-E'_1\right)             \label{W1ex3}\\
   {} [W_{1}]&= i_{R*}\left(3l'-E'_1-\dots-E'_7\right) \label{W1ex4}
\end{align}
\end{subequations}

We can analyze the moduli spaces in the $dP_9$ surface of each of
these different cases by considering the right-hand sides
of~\eqref{W1ex} as classes of curves through some number of points in
$\PP^2$. In examples \eqref{W1ex1} and \eqref{W1ex2}, the curves are
always reducible to just two copies of an exceptional curve in $dP_8$
and, hence, have no moduli for moving in $dP_{9}$. Since the curve $R$
is also fixed in the base, these cases have no moduli for moving in
the Calabi--Yau manifold at all. 

The third case, \eqref{W1ex3}, corresponds to a line (topologically
$\PP^1$) in $\PP^2$ through one point. As such, as discussed in
section~\ref{exbase1}, its moduli space is $\PP^1$. Furthermore, there
are special points in the moduli space where the line passes through
one of the other blown-up points in $dP_9$ and the curve becomes
reducible (in analogy to the process described in
Figure~\ref{lE1}). We can have, for instance,
\begin{equation}
   W_1 = U_1 + U_2
\label{W1split}
\end{equation}
where the classes in $dP_9$ are
\begin{equation}
   [U_1]_R = l' - E'_1 - E'_2 \qquad
   [U_2]_R = E'_2
\end{equation}
What was previously a single sphere has now been reduced to a pair of
spheres, each an exceptional curve, which intersect at a single
point (as in Figure~\ref{Cpinch}). 

\begin{figure}
   \centerline{\psfig{figure=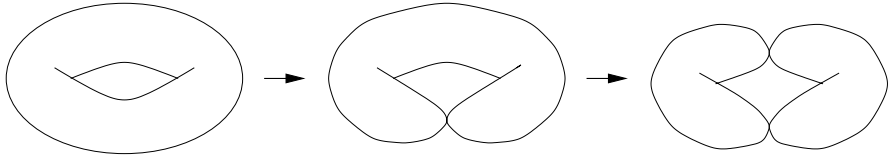,height=0.8in}}
   \caption{Pinching a torus at one point and at a pair of points}
   \label{toruspinch}
\end{figure}

The last case, \eqref{W1ex4}, is even more interesting. This class
corresponds to a cubic in $\PP^2$ passing through 7 points. A cubic is an
elliptic curve and so has genus one. A general cubic has a moduli space of
$\PP^9$. However, by being restricted to pass through seven points, the
remaining moduli space is simply $\PP^2$. At special points in the
moduli space the cubic can degenerate. First, we can have a double
point. This occurs when the discriminant of the curve vanishes. It
corresponds to one of the cycles of the torus pinching, as shown in
Figure~\ref{toruspinch}. The vanishing of the discriminant is a single
additional condition on the parameters and so gives a curve, which we
will call $\D_{W_1}$, in $\PP^2$. When the cubic degenerates, the
blown up curve is a sphere. Thus, it has changed genus. We can now go
one step further. At certain places, the discriminant curve $\D_{W_1}$
in $\PP^2$ has a double point. This corresponds to places where the
curve becomes reducible. The curve $W_1$ splits into two
\begin{equation}
    W_1 = U_1 + U_2
 \label{eq:today2}
\end{equation}
with, for instance, $U_1$ and $U_2$ describing a line and a conic, 
\begin{equation}
   [U_1]_R = l' - E'_1 - E'_2 \qquad
   [U_2]_R = 2l' - E'_3 - \dots - E'_7
\end{equation}
Note that each of the resulting curves is exceptional and so has no
moduli space in the $dP_9$. However, we note that this
splitting can happen $\binom{7}{2}=21$ different ways. The two curves
intersect at two points, corresponding to a double pinching of the
torus into a pair of spheres, as in Figure~\ref{toruspinch}. For
completeness, we note that there is one further singularity
possible. That is where the cubic develops a cusp. However, the
topology remains that of a sphere, so we will not distinguish these
points. 

We can summarize these branches of moduli space in the following table
\begin{equation}
\begin{array}{c|cc}
   [W_1] \text{ in $dP_9$} & \text{genus} & \text{moduli space} \\
   \hline \hline
   2E'_1  & 
      0+0 & \PPa\ts\PPa/\ZZ_2 \\
   \hline
   E'_1+E'_2  &
      0+0 & \PPa \ts \PPa \\
   \hline
   l'-E'_1  &
      0 & \left(\PP^1-8\text{ pts.}\right) \ts \PPa \\
   \left(l'-E'_1-E'_2\right) + \left(E'_2\right) &
      0+0 & \PPa \ts \PPa \\
   \hline
   3l'-E'_1-\dots-E'_7  &
      1 & \left(\PP^2-\D_{W_1}\right) \ts \PPa \\
   3l'-E'_1-\dots-E'_7  &
      0 & \left(\D_{W_1}-21\text{ pts.}\right) \ts \PPa \\
   \left(l'-E'_1-E'_2\right)+\left(2l'-E'_3-\dots-E'_7\right) &
      0+0 & \PPa \ts \PPa 
\end{array}
\label{W1mod}
\end{equation}
Note that in the first line, we mod out by $\ZZ_2$ since the fivebranes
wrap the same curve in the Calabi--Yau manifold and so are
indistinguishable. This is not the case in the second example.  

We note that there is another type of splitting possible for the fourth
example. The cubic could pass though one of the additional blown up
points. It would then pass through eight of the nine points. However,
as we have discussed above, it must then also pass through the ninth
point. The curve $W_1$ would then decompose into a pure fiber
component plus two sections, namely
\begin{equation}
   W_1 = U_1 + U_2 + U_3
\end{equation}
with 
\begin{equation}
   [U_1]_R = F_R = 3l'-E'_1-\dots-E'_9, \qquad
   [U_2]_R = E'_8, \qquad
   [U_3]_R = E'_9
\end{equation}
Thus we see that, unlike the case for $W_2$ and $W_3$, it is also
possible to have a transition where a fiber component splits off from
$W_1$.

Let us end this section with an important observation. 
We have not discussed the full moduli space of $W$ here but only
various characteristic branches. There is, however, a certain type of
branch that we would like to emphasize. Consider a branch where we
choose $n=17$, so that $W_F=0$, and split the curve $W$ as follows 
\begin{equation}
   W = W_0 = W_1 + W_2 + W_3
\end{equation}
with, for example
\begin{equation}
   [W_1] = 2\s_*E_1 + 2F, \qquad
   [W_2] = \s_*E_2 + 5F, \qquad
   [W_3] = \s_*E_3 + 10F
\end{equation}
From our previous discussion, $W_2$ and $W_3$ are required to be
sections of the $dP_9$ surface above the exceptional curve in the
class $E_2$ and $E_3$ respectively. As such, they have no moduli to
move within the Calabi--Yau threefold. Furthermore, we have seen an
example in the second line of the table~\eqref{W1mod} where $W_1$
splits into two components, neither of which can move in the
Calabi--Yau space. Consequently, we see that there is a component of
moduli space where we simply have four fivebranes, each wrapping a fixed
curve within the Calabi--Yau threefold. Furthermore, none of the
fivebranes can intersect. We then have a very simple moduli space. It
is 
\begin{equation}
   \PPa \ts \PPa \ts \PPa \ts \PPa 
\end{equation}
corresponding to moving each fivebrane in $\orb$ and changing the
value of the axions. Thus, even when the fivebrane class is relatively
complicated, we see that there are components of the moduli space with
very few moduli.


\section{A second three-family $dP_8$ example}

In this section, we will briefly discuss a second example of a
realistic fivebrane moduli space. Again, we will take the base
\begin{equation}
   B \text{ is a $dP_8$ surface}
\end{equation}
and choose the fivebrane class
\begin{equation}
   [W] = \s_*l - \s_*E_1 + \s_*E_2 + \s_*E_3 + 27F
\end{equation}
This gives a three family model with an unbroken $SU(5)$ gauge
group, as can be seen explicitly using the rules given in~\cite{dlow1}
and~\cite{dlow2} (taking $\l=\frac{1}{2}$ in the equations given
there). From the condition~\eqref{thrm}, since $l-E_1+E_2+E_3$ is
effective in the base, $[W]$ is an effective class as required. 

To calculate the moduli space, first, one separates the pure fiber
components, partitioning $W$ as $W=W_0+W_F$, as in~\eqref{decompF},
with
\begin{equation}
   [W_0] = \s_*l - \s_*E_1 + \s_*E_2 + \s_*E_3 + nF, \qquad
   [W_F] = \left(27-n\right) F
\end{equation}
where $0\leq n\leq 27$. As usual, unless $n=27$, this implies that we
have at least two distinct fivebranes. This partition splits the
moduli space into 28 components, as in~\eqref{moddecomp}. The moduli
space of the $W_F$ component is the usual symmetric
product given in~\eqref{WFmod}
\begin{equation}
   \cM((27-n)F) = \left( F_r \ts \PPa\right)^{27-n}/\ZZ_{27-n}
\label{WFex2}
\end{equation}

Next, we analyze the moduli space of $W_0$ for a given $n$. If we
project $W_0$ onto the base, we get the curve $C$ in the class
\begin{equation}
   [C] = l - E_1 + E_2 + E_3
\end{equation}
We then need to find the moduli space $\cM_B(l-E_1+E_2+E_3)$ of such
curves in the base. The last two classes correspond to curves wrapping
exceptional blow-ups and so, as in the previous section, $C$ must be
reducible. We expect that there is always one component wrapping the
exceptional curve in $E_2$ and another component wrapping the
exceptional curve in $E_3$. 

In general, there are eight distinct parts of $\cM_B(l-E_1+E_2+E_3)$,
with different numbers of curves in the base. In general, $C$
decomposes into $k$ curves as 
\begin{equation}
   C = C_1 + \dots + C_k
\end{equation}
with
\begin{equation}
  [C_i] = \O_i, \qquad \O_1 + \dots + \O_k = l - E_1 + E_2 + E_3
\end{equation}
As $C$ splits, so does $W_0$. In general we can partition the $n$
fiber components in different ways, so that 
\begin{equation}
   W = W_1 + \dots + W_k, \qquad [W_i] = \s_*\O_i + n_iF
\end{equation}
where $n_i\geq 0$ and $n_1+\dots+n_k=n$. The eight different parts of
the moduli space of $C$ can be summarized as follows 
\begin{equation}
\begin{array}{cc|c}
   [C_i]    & \genus{C_i} & \text{moduli space} \\
   \hline\hline
     \begin{array}{c}
        {}[C_1] = l - E_1 \\ 
        {}[C_2] = E_2 \\ 
        {}[C_3] = E_3 
     \end{array}
     & 
     \begin{array}{c} 0 \\ 0 \\ 0 \end{array}
     & \PP^1 - 7\text{ pts.}
   \\ \hline
     \begin{array}{c}
        {}[C_1] = l - E_1 - E_2 \\ 
        {}[C_2] = 2E_2 \\ 
        {}[C_3] = E_3 
     \end{array}
     & 
     \begin{array}{c} 0 \\ 0 \\ 0 \end{array}
     & \text{single pt.}
   \\ \hline
     \begin{array}{c}
        {}[C_1] = l - E_1 - E_3 \\ 
        {}[C_2] = E_2 \\ 
        {}[C_3] = 2E_3 
     \end{array}
     & 
     \begin{array}{c} 0 \\ 0 \\ 0 \end{array}
     & \text{single pt.}
   \\ \hline
     \begin{array}{c}
        {}[C_1] = l - E_1 - E_i \\ 
        {}[C_2] = E_2 \\ 
        {}[C_3] = E_3 \\ 
        {}[C_4] = E_i
     \end{array}
     & 
     \begin{array}{c} 0 \\ 0 \\ 0 \\ 0 \end{array}
     & 
     \begin{array}{c} \text{single pt.} \\ i=4,\dots,8 \end{array}
\end{array}
\label{modCtable}
\end{equation}
The analysis is very similar to that in section~\ref{exbase1}. The
$C_2$ and $C_3$ components are always stuck on the exceptional curves
in $E_2$ and $E_3$ and so have no moduli. In the first row in the
table, $C_1$ is a curve in the $dP_8$ corresponding to a line through
the blown-up point $p_1$. As such it has a moduli of $\PP^1$, except
for the seven special points where it also passes through one of the
other seven blown-up points. If it passes through one of the points
$p_4,\dots,p_8$, the curve splits into two curves and we have a total
of four distinct fivebranes. This case is given in the last row 
in~\eqref{modCtable}. If it passes through $p_2$ or $p_3$, generically,
we still only have three components, but now $C_2$ or $C_3$ is in the
class $2E_2$ or $2E_3$ respectively. These are the second and third
rows in~\eqref{modCtable}. All the curves are spheres so have genus
zero. In conclusion, we have
\begin{equation}
   \cM_B(l-E_1+E_2+E_3) = \PP^1
\end{equation}
Where there are seven special points in the moduli space: the five
cases given in the last row in~\eqref{modCtable}, where $C$ splits
into four curves, and the two further points where the class of $C_2$
and $C_3$ changes as given in the second and third rows
of~\eqref{modCtable}. We see that, unlike the previous example,
$\cM_B$ is more than just a single point.

Let us concentrate on a generic point in moduli space, as in the first
row of~\eqref{modCtable}. As we noted above, $W_0$ then splits into
three distinct components, 
\begin{equation}
   W_0 = W_1 + W_2 + W_3
\end{equation}
with 
\begin{equation}
   [W_1] = \s_*l - \s_*E_1 + n_1 F, \qquad
   [W_2] = \s_*E_2 + n_2 F, \qquad
   [W_3] = \s_*E_3 + n_3 F
\label{ex2split1}
\end{equation}
where $n_1+n_2+n_3=n$ and $n_i\geq 0$. 

We note that $W_2$ and $W_3$ are exactly of the form we analyzed in
section~\ref{W2W3} above. Recall that each component is a curve stuck
above the exceptional curve $C_2$ or $C_3$ in the base. The space
$S_{C_i}=\pi^{-1}(C_i)$, for $i=2,3$, above each exceptional curve was 
given by 
\begin{equation}
   S_{C_i} \text{ is a $dP_9$ surface for $i=2,3$}
\end{equation}
This means we have $g_i=0$ and $p_i=1$ and the corresponding moduli
spaces are given by $\cM(0,1;n_i)$. These spaces were given
in~\eqref{cMm} and are a discrete number $N(n_i)$ of copies of $\PPa$,
corresponding to different sections of a $dP_9$ surface. We have 
\begin{equation}
   \cM(0,1;n_i) = N(n_i) \PPa \qquad \text{for $i=2,3$}
\label{modW2W3}
\end{equation}
There are no moduli for moving the curves within either the base or
the fiber of the Calabi--Yau manifold. Since $g_i=0$ there are no
vector multiplets on the fivebranes. 

We now turn to $W_1$. This is essentially of the form we considered
section~\ref{exbase2}. There, we showed that the surface $S_{C_1}$ is
given by
\begin{equation}
   S_{C_1} \text{ is a K3 surface}
\end{equation}
That is to say, the genus $g_1$ of $C_1$ is zero and $p_1=2$, so there
are 24 singular fibers. Thus we are interested in the moduli space
$\cM(0,2;n_1)$. From the general discussion in section~\ref{generic},
we find from equation~\eqref{gencMR2} that the moduli space for $W_1$
is empty unless $n_1=0$. One then has 
\begin{equation}
   \cM(0,2;0) = \PPa
\end{equation}
Putting this together with the moduli space for $W_2$ and $W_3$ given
in~\eqref{modW2W3}, and recalling the form of the $C$ moduli
space~\eqref{modCtable}, we can write, for this generic part of the
moduli space of $W_0$, 
\begin{multline}
   \cM_0(\s_*l-\s_*E_1+\s_*E_2+\s_*E_3+nF)  \\
        = \left(\left[\PP^1 - 7\text{ pts.}\right]\ts\PPa\right) 
             \ts N(n_2)\PPa \ts N(n_3)\PPa
          + \cM_{\text{non-generic}}
\end{multline}
where we must have $n_1=0$, and $n_2+n_3=n$ in the
decomposition~\eqref{ex2split1}. We have three distinct fivebranes,
two of which, $W_2$ and $W_3$ are stuck in the Calabi--Yau
threefold. The third fivebrane can move within the base of the
Calabi--Yau. Each fivebrane wraps a curve of genus zero and so there
are no vector multiplets. The full $W$ moduli space is constructed,
for each $n$, from the product of this space together with the $W_B$
moduli space given in~\eqref{WFex2}. 

We will not consider all the seven possible exceptional cases in the
moduli space of $C$, listed in~\eqref{modCtable}. Rather, consider
just one of the cases where $C$ splits into four components,
\begin{equation}
   [C_1] = l - E_1 - E_4, \qquad
   [C_2] = E_2, \qquad
   [C_3] = E_3, \qquad
   [C_4] = E_4
\end{equation}
with the corresponding split of $W_0$ as
\begin{equation}
\begin{aligned}
   {}[W_1] &= \s_*l - \s_*E_1 - \s_*E_4 + n_1 F, \quad &
   {}[W_2] &= \s_*E_2 + n_2 F, \\
   {}[W_3] &= \s_*E_3 + n_3 F, &
   {}[W_4] &= \s_*E_4 + n_4 F
\end{aligned}
\end{equation}
with $n_i\geq 0$ and $n_1+n_2+n_3+n_4=n$. Let us further assume that
$n=27$ so that $W_F=0$ and there are no pure fiber components. Each of
the $C_i$ is an exceptional curve in the base. Consequently, we have 
\begin{equation}
   S_{C_i} \text{ is a $dP_9$ surface for $i=1,2,3,4$}
\end{equation}
(For $C_1$ the calculation is just as in section~\ref{exbase2}). Thus
we have $g_i=0$ and $p_i=1$ for each curve and so we have the moduli
spaces
\begin{equation}
   \cM(0,1;n_i) = N(n_i)\PPa \qquad \text{for } i = 1,2,3,4
\end{equation}
where, as usual, $N(n_i)$ counts the number of distinct sections in
each $dP_9$. In particular, we are no longer required to take
$n_1=0$. We now have a total of four distinct fivebranes, wrapping
$W_1$, $W_2$, $W_3$ and $W_4$. All these curves are stuck in the
Calabi--Yau threefold, so that a given connected part of the moduli
space has the form    
\begin{equation}
   \PPa \ts \PPa \ts \PPa \ts \PPa
\end{equation}
(Here we have assumed that either $n_1$ or $n_4$ is non-zero so that
this branch of moduli space if not connected to the generic branch
discussed above.) As in the previous section, we see that there are
disconnected components of the moduli space with very few moduli. Each
curve has genus zero, so there are no vector multiplet degrees of
freedom.


\section{Two simple Hirzebruch examples}

In this section, we will briefly discuss two simple examples where the
base is a Hirzebruch surface $F_r$. These are $\PP^1$ fibrations over
$\PP^1$, characterized by a non-negative integer $r$. Following the
notation of~\cite{dlow2}, they have two independent algebraic
classes, the class $\cS$ of the section of the fibration at infinity
and the fiber class $\cE$. These have the following intersection
numbers 
\begin{equation}
   \cS \cdot \cS = -r, \qquad
   \cS \cdot \cE = 1, \qquad
   \cE \cdot \cE = 0
\end{equation}
The canonical bundle is given by 
\begin{equation}
   K_B = -2 \cS - (2+r) \cE
\end{equation}
Effective classes in $F_r$ are of the form $\O=a\cS+b\cE$ with $a$ and
$b$ non-negative. Thus, from~\eqref{thrm}, a general class of effective
curves in the Calabi-Yau threefold can be written as
\begin{equation}
   [W] = a \s_*\cS + b \s_*\cE + f F
\end{equation}
where $a$, $b$ and $f$ are all non-negative. (Note, however, that, as
discussed above equation~\eqref{thrm}, there is actually an additional
effective class for $r\geq 3$, which we will ignore here.) 

For any realistic model with three families of matter and realistic
gauge groups, if the hidden $E_8$ group is unbroken, then the
coefficients $a$ and $b$ are typically large~\cite{dlow2}. The moduli
space is then relatively complicated to analyze. Thus, for simplicity,
we will consider only two very simple cases with either a single $\cS$
class or a single $\cE$ class.

\subsection{$[W]=\s_*\cS+fF$}

We consider first
\begin{equation}
   B \text{ is an $F_r$ surface with $r\geq 2$}
\end{equation}
and
\begin{equation}
   [W] = \s_*\cS + f F
\end{equation}
using the general procedure we outlined above. Note that we require
$r\geq 2$ to exclude the trivial case of $F_0$, which is just the
product $\PP^1\ts\PP^1$, and $F_1$, which is actually the del Pezzo
surface $dP_1$. 

We recall that the first step is to split off any pure fiber components
from $W$, writing it as a sum of $W_0$ and $W_F$ as in~\eqref{decompF}
\begin{equation}
   [W_0] = \s_*\cS + n F, \qquad
   [W_F] = (f-n) F 
\label{cSpartition}
\end{equation}
with $0\leq n \leq f$. As usual, unless $n=f$, this implies we have at
least two distinct fivebranes. This splits the moduli space into $f+1$
components as in~\eqref{moddecomp}. The moduli space of $W_F$ has the
familiar form, from~\eqref{WFmod}, 
\begin{equation}
   \cM((f-n)F) = \left( F_r \ts \PPa\right)^{f-n}/\ZZ_{f-n}
\end{equation}

Next we turn to analyzing the $W_0$ moduli space for a given
$n$. Projecting $W_0$ onto the base gives the curve $C$ in the
homology class
\begin{equation}
   [C] = \cS
\end{equation}
Our first step is then to find the moduli space $\cM_B(\cS)$ of $C$ in
the base. This, however, is very simple. Since the self-intersection of
$\cS$ is negative, there is a unique representative of the class $\cS$,
namely the section at infinity. Thus 
\begin{equation}
   \cM_B(\cS) = \text{single point}
\end{equation}
The second step is the to find all the curves $W_0$ in the class
$\p_*\cS+nF$ in the Calabi--Yau threefold which project onto $C$. To
answer this, we need to characterize the surface $S_C=\p^{-1}(C)$. We
recall that this will be an elliptic fibration over $C$ and is
characterized by the genus $g$ of $C$ and the number $12p$ of singular
fibers. Since $C$ is a section of the Hirzebruch surface, it must be
topologically $\PP^1$. Consequently
\begin{equation}
   g = \text{genus}(C) = 0
\label{genuscS}
\end{equation}
The expected number number of singular fibers is given by $12p$, where
\begin{equation}
\begin{aligned}
   p &= - K_B \cdot [C] \\
     &= \left(2\cS + (2+r)\cE\right) \cdot \cS = 2 - r 
\end{aligned}
\label{pcS}
\end{equation}
For $r\geq 3$ this seems to predict that we have a negative number of
singular fibers. Actually, this reflects the fact that the curve $C$ is
contained within the discriminant curve of the Calabi--Yau elliptic
fibration. Thus the elliptic curve over $C$ is singular
everywhere. This is a non-generic case we wish to avoid in discussing
the moduli space. Thus, we will assume $r<3$. Given that $F_0$ and
$F_1$ do not give new surfaces, we are left with restricting to $r=2$
and so 
\begin{equation}
   B \text{ is $F_2$}
\end{equation}
Then we have $p=0$ which implies the elliptic fibration is locally
trivial. If we assume that the surface $S_C$ is also a globally
trivial fibration, it is then simply the product 
\begin{equation}
   S_C = \PP^1 \ts E 
\end{equation}
where $E$ is an elliptic curve. Comparing with the notation of
section~\ref{gen}, we have $g=p=0$ and so we are interested in the
moduli space $\cM(0,0;n)$. However, we have argued,~\eqref{gencMR2},
that these spaces are empty unless $n=0$. In this case
$\cM(0,0;0)=E\ts\PPa$, where the first factor comes from moving the
fivebrane within $\orb$ and the second factor comes from moving it
within $S_C$. (Again, here we are assuming that $S_C$ is a globally as
well as locally trivial fibration.) In particular, the curve is a
section of $S_C$ and, so, can lie at any point on the elliptic curve
$E$. 

Thus, we see that the only partition~\eqref{cSpartition} allowed is
where $n=0$. In that case the moduli space of the curve $W_0$ is
simply
\begin{equation}
   \cM_0(\s_*\cS) = E \ts \PPa
\end{equation}
The full moduli space is then the product
\begin{equation}
   \cM(\s_*\cS+fF) = \left(E \ts \PPa\right) 
        \ts \frac{\left( F_2 \ts \PPa\right)^f}{\ZZ_f}
\end{equation}
Generically, we have $f+1$ fivebranes. One is $W_0$, which lies over the
section at infinity in $F_2$ and can be at any position in the
elliptic fiber over the base as well as in $\orb$. The other $f$
fivebranes are pure fiber components, which can each be at arbitrary
points in the base and in $\orb$. Since the genus of $C$ is zero, so
is the genus of $W_0$. Thus, generically, the only vector multiplets
come from $W_F$. At a generic point in moduli space, we therefore have $U(1)^f$
gauge symmetry.  It is possible for the fiber components to intersect
$W_0$, in which case we might expect new massless fields to appear in
the low-energy theory.

\subsection{$[W]=\s_*\cE+fF$}

Now consider the case where again
\begin{equation}
   B \text{ is a $F_r$ surface with $r\geq 2$}
\end{equation}
but we take 
\begin{equation}
   [W] = \s_*\cE + f F
\end{equation}
As always, we first split off the pure fiber components in $W$, writing
$W=W_0+W_F$ with $f+1$ different partitions
\begin{equation}
   [W_0] = \s_*\cS + n F, \qquad
   [W_F] = (f-n) F 
\label{cEpartition}
\end{equation}
The moduli space of $W_F$ is the familiar form
\begin{equation}
   \cM((f-n)F) = \left( F_r \ts \PPa\right)^{f-n}/\ZZ_{f-n}
\end{equation}

To analyze the the moduli space of $W_0$, we project onto the base,
giving the curve $C$ with 
\begin{equation}
   [C] = \cE
\end{equation}
Thus $C$ is in the fiber class of the $F_n$ base. Since $F_n$ is a
$\PP^1$ fibration over $\PP^1$, the fiber can lie at any point in the
base $\PP^1$ so we have that the moduli space of $C$ is given by 
\begin{equation}
   \cM_B(\cE) = \PP^1
\label{cEbase}
\end{equation}
Next we need to find all the curves $W_0$ which project onto a given
$C$. We first characterize the surface $S_C=\p^{-1}(C)$. Since $C$ is
a fiber of $F_r$, it must be topologically $\PP^1$. Hence
\begin{equation}
   g = \text{genus}(C) = 0 
\end{equation}
The number of singular fibers, $12p$, in the elliptic fibration $S_C$
given in terms of 
\begin{equation}
\begin{aligned}
   p &= -K_B \cdot [C] \\
     &= \left(2\cS+(2+r)\cE\right) \cdot \cE = 2
\end{aligned}
\end{equation}
for any $r$. This implies that $S_C$ is an elliptic fibration over
$\PP^1$ with 24 singular fibers, that is 
\begin{equation}
   S_C \text{ is a K3 surface}
\end{equation}
In the notation of section~\ref{gen}, we have $g=0$ and $p=2$ and so
we are interested in the moduli space $\cM(0,2;n)$. However, we have
argued,~\eqref{gencMR2}, that these spaces are empty unless $n=0$, in
which case $\cM(0,2;0)=\PPa$. That is, the curve is completely stuck
within $S_C$, but is free to move within $\orb$. 

Thus the only partition~\eqref{cEpartition} allowed is where $n=0$. In
this case, the full moduli space of $W_0$ is then given by
\begin{equation}
   \cM_0(\s_*\cE) = \PP^1 \ts \PPa
\end{equation}
The first factor of $\PP^1$ reflects the moduli space of curves $C$ in
the base~\eqref{cEbase}. The second factor reflects the fact that, for
a given $C$, the moduli space of $W_0$ is $\cM(0,2;0)=\PPa$. We note
that there are no moduli for moving $W_0$ in the direction of the
elliptic fiber. The full moduli space is then given by the product 
\begin{equation}
   \cM(\s_*\cE+fF) = \left(\PP^1 \ts \PPa\right) 
        \ts \frac{\left( F_r \ts \PPa\right)^f}{\ZZ_f}
\end{equation}
As in the previous example, generically we have $f+1$ fivebranes. One is
$W_0$ which, in the base, can be deformed in the base and can move in
$\orb$, but has no moduli for moving in the elliptic fiber. The other
$f$ fivebranes are pure fiber components, which can each be at
arbitrary points in the base and in $\orb$. Again, since the genus of
$C$ is zero, so is the genus of $W_0$. Thus, generically, the only
vector multiplets come from $W_F$. At a generic point in moduli space
we have, therefore, $U(1)^f$ gauge symmetry.  It is possible for the fiber
components to intersect $W_0$, in which case we might expect new
massless fields to appear in the low-energy theory. 

\subsection*{Acknowledgments}

B.A.O. and D.W. would like to thank Angel Uranga for helpful
discussions. 
R.D. is supported in part by an NSF grant DMS-9802456 as well as by
grants from the University of Pennsylvania Research Foundation and
Hebrew University. 
B.A.O. is supported in part by a Senior Alexander von Humboldt Award, 
by the DOE under contract No. DE-AC02-76-ER-03071 and by a University 
of Pennsylvania Research Foundation Grant. 
D.W. is supported in part by the DOE under contract
No. DE-FG02-91ER40671.



\end{document}